\documentclass[%
 reprint,
 amsmath,amssymb,
 aps,
]{revtex4-2}
\usepackage{graphicx}% Include figure files
\usepackage{dcolumn}% Align table columns on decimal point
\usepackage{bm}% bold math
\usepackage{xcolor}
\usepackage{siunitx}

\DeclareSIUnit\ppm{ppm} 
\DeclareSIUnit\sqrthz{\ensuremath{\sqrt{\text{Hz}}}}
\DeclareSIUnit[per-mode = symbol]\Hzasd{\Hz\per\sqrthz} 
\DeclareSIUnit[per-mode = symbol]\masd{\m\per\sqrthz} 
\DeclareSIUnit[per-mode = symbol]\pmasd{\pico\m\per\sqrthz}
\DeclareSIUnit[per-mode = symbol]\umasd{\micro\m\per\sqrthz}

\newcommand{\B}[1]{\textcolor{blue}{#1}}

\begin{document}

\preprint{APS/123-QED}

\title{First result for testing semiclassical gravity effect with a torsion balance}

\author{Tianliang Yan}
  \email{tianliang.yan01@gmail.com}
  \affiliation{University of Birmingham, School of Physics and Astronomy, Birmingham B15 2TT, United Kingdom.}
\author{Yubao Liu}
\email{lybphy@hust.edu.cn}
\affiliation{National Gravitation Laboratory, Hubei Key Laboratory of Gravitation and Quantum Physics, School of Physics, Huazhong University of Science and Technology, Wuhan 430074, China.}%
\author{Leonid Prokhorov}
\affiliation{University of Birmingham, School of Physics and Astronomy, Birmingham B15 2TT, United Kingdom.}
\author{Jiri Smetana}
\affiliation{University of Birmingham, School of Physics and Astronomy, Birmingham B15 2TT, United Kingdom.}
\author{Haixing Miao}
\affiliation{Frontier Science Center for Quantum Information, Department of Physics, Tsinghua University, Beijing 100084, China}
\author{Yiqiu Ma}
\affiliation{National Gravitation Laboratory, Hubei Key Laboratory of Gravitation and Quantum Physics, School of Physics, Huazhong University of Science and Technology, Wuhan 430074, China.}%
\author{Vincent Boyer}
\affiliation{University of Birmingham, School of Physics and Astronomy, Birmingham B15 2TT, United Kingdom.}
\author{Denis Martynov}
\affiliation{University of Birmingham, School of Physics and Astronomy, Birmingham B15 2TT, United Kingdom.}%

\date{\today}

\begin{abstract}
The Schrödinger-Newton equation, a theoretical framework connecting quantum mechanics with classical gravity, predicts that gravity may induce measurable deviations in low-frequency mechanical systems—an intriguing hypothesis at the frontier of fundamental physics. In this study, we developed and operated an advanced optomechanical platform to investigate these effects. The system integrates an optical cavity with finesse over \SI{3.5e5}{} and a torsion pendulum with an ultra-low eigenfrequency of \SI{0.6}{\mHz}, achieving a high mechanical quality factor exceeding $5\times10^4$. We collected data for 3 months and reached a sensitivity of \SI{0.3}{\micro rad/\sqrthz} at the Schr\"{o}dinger-Newton frequency of \SI{2.5}{\mHz} where deviations from the standard quantum mechanics may occur. While no evidence supporting semiclassical gravity was found, we identify key challenges in such tests and propose new experimental approaches to advance this line of inquiry. This work demonstrates the potential of precision optomechanics to probe the interplay between quantum mechanics and gravity.

\end{abstract}

\maketitle
\section{\label{sec:intro}Introduction}
The semi-classical gravity theory relates classical Einstein tensors with the quantum expectation value of the energy-momentum tensor of quantum matter\,\cite{Rosenfeld1963,Mueller1962}, as given by:
\begin{equation}
G_{\rm \mu\nu}=\frac{8\pi G}{c^4}\langle\hat{T}_{\rm \mu\nu}\rangle.
\end{equation}
Despite its drawbacks, such as issues with divergence-free conditions and potential superluminal communication\,\cite{Simon_2001,Mielnik_20011}, semi-classical gravity has not been ruled out experimentally. Therefore, testing this theory remains important for understanding the nature of spacetime.

Recent developments in high-precision measurement allow us to use the optomechanical platform to test semi-classical gravity\,\cite{Jayich2008,Chen_2013,Aspelmyer2014,Hoang2016,Rossi2018,Mason2019,Yu2020,Aggarwal2020,Cata2020}. As theoretical analysis suggests, the key to distinguishing between quantum and SN gravity in an optomechanical system is the preparation of the test mass with a very low eigenfrequency. In this paper, we experimentally focus on the detection of the signals of semiclassical gravity.
We focus on a simple optomechanics setup, in which a torsion pendulum is monitored by laser light. Taking the effect of SN gravity into consideration, the evolution of the system is dominated by the following Hamiltonian \cite{Helou2017}:
\begin{equation}\label{eq:Hamiltonian_self_gravity}
\hat{H}=\frac{\hat{p}^2}{2M}+\frac{1}{2}M\omega_{\rm m}^2\hat{x}^2+\frac{1}{2}M\omega_{\rm{SN}}^2(\hat{x}-\langle\hat{x}\rangle)^2-\hbar\alpha\hat{a}_1\hat{x},
\end{equation}
where $M$ is the mass of the test mass, \(\hat{p}\) is the center-of-mass momentum operator, $\omega_{\rm m}$ is the eigenfrequency of the test mass, \(\hat{x}\) is the center-of-mass position operator, $\alpha$ is the coupling strength of the test mass and laser light and $\hat a_1$ is the amplitude operator of the optical field. $\omega_{\rm SN}$ is a frequency scale determined by the object's matter distribution. For materials with single atoms sitting close to lattice sites, we have\,\cite{yang2013macroscopic}:
\begin{equation}
 \omega_{\rm SN} = \frac{Gm}{6\sqrt{\pi} \Delta x_{\rm int}^3}
 \label{eq:SN f},
\end{equation}
in which $G$ denotes the gravitational constant, while $m$ refers to the atomic mass. The term $\Delta x_{\rm int}$ is defined as the standard deviation of the internal displacement of a test mass's constituent atoms from their equilibrium positions in every spatial direction, attributed to quantum fluctuations. Typically, for an aluminum mass, $\omega_{\rm SN} = 2 \pi \times \SI{2.51}{\mHz}$. 

Assuming a Gaussian initial mechanical state, Yang et al. \cite{yang2013macroscopic} demonstrate that the signature of Eq.\,(\ref{eq:Hamiltonian_self_gravity}) manifests in the rotation of the Wigner function of the mechanical state, with frequency
\begin{equation}
\omega_{\rm q} = \sqrt{\omega_{\rm m}^2 + \omega_{\rm SN}^2}
\label{eq:omega_q}.
\end{equation}
Drawing from Eq.~(\ref{eq:omega_q}), \(\omega_{\rm q}\) becomes harder to distinguish from \(\omega_{\rm m}\) as \(\omega_{\rm m}\) increases. Therefore, it is essential to ensure that $\omega_{\rm m}\sim\omega_{\rm SN}$ to maximize the difference. Moreover, Helou\,\emph{et al.} show that the optomechanical system under the influence of the quadratic SN self-gravity potential could have different responses to classical noise and quantum noise, leaving a double-peak structure in the spectrum of the light field reflecting from the test mass, provided that the pre-selection prescription of quantum measurement is assumed\,\cite{Helou2017,Liu2023}. These two peaks locate at $\omega_m$ and $\omega_q$ at the outgoing field spectrum, which corresponds to the classical and quantum noise response, respectively. It is this double-peak structure of the outgoing optical spectrum that is the signal that we are targeting in the experiment.

In Birmingham, we have developed a six-degree-of-freedom (6D) seismometer\,\cite{mow20196d, ubhi2022active, prokhorov2023design} with a torsional (RZ) resonant frequency of \SI{0.6}{\mHz}. 
Previous studies have also explored the integration of a torsion pendulum with an optical cavity for precision measurements\,\cite{komori2020attonewton, chua2023torsion}.
By integrating a cavity into this seismometer, we constructed an optomechanical system with a test mass frequency that meets the required specifications.

In this paper, we present a low-mechanical-frequency optomechanical system designed to test the semiclassical gravity effect. This system includes a cavity with a finesse over \SI{3.5e5}{}, with the output coupler mounted on a torsion pendulum featuring a \SI{0.6}{\mHz} torsional eigenmode. The high finesse amplifies the quantum radiation pressure effect to excite the SN signal. We estimate the pendulum's $Q$-factor to be at least $5 \times 10^4$, which optimizes the signal-to-noise ratio (SNR) for detection. To maintain cavity lock stability and minimize noise, we employed two seismic noise suppression servos, two laser frequency stabilization servos, one servo and one feed-forward for classical radiation pressure noise suppression, along with one laser intensity stabilization servo. We achieved a sensitivity of \SI{0.3}{\micro rad/\sqrthz} at \SI{2.5}{\mHz} where deviations from the standard quantum mechanics may occur. Based on these experimental results, we introduce two potential enhancement methods geared toward improving the detection of the SN signature. These upgrade approaches will specifically target the examination of the SN effect within both the quadratic and non-quadratic regions of the SN self-gravity potential, thereby enriching our comprehension of the SN phenomenology. Furthermore, this versatile system promises to be a valuable tool for exploring other physical phenomena at the intersection of quantum mechanics and general relativity \cite{miao2020quantum}.

%this adaptable system is anticipated to be instrumental in investigating other physical phenomena where quantum mechanics and general relativity intersect\,\cite{miao2020quantum}.

\section{Experimental Layout}
\subsection{The optical setup}
The six-degree-of-freedom (6D) seismometer\,\cite{ubhi2022active, prokhorov2023design} consists of a mass suspended by a single fused-silica fiber with a length of \SI{55}{cm} and a diameter of \SI{100}{\micro m}. It surpasses the sensitivity of existing commercial seismometers by over an order of magnitude in the angular degrees of freedom. Also, it has a very low torsion mode of \SI{0.6}{\mHz}. These two factors make the seismometer suitable for the test of semiclassical gravity effect.

\begin{figure*}
\centering
\includegraphics[width=1.0\textwidth]{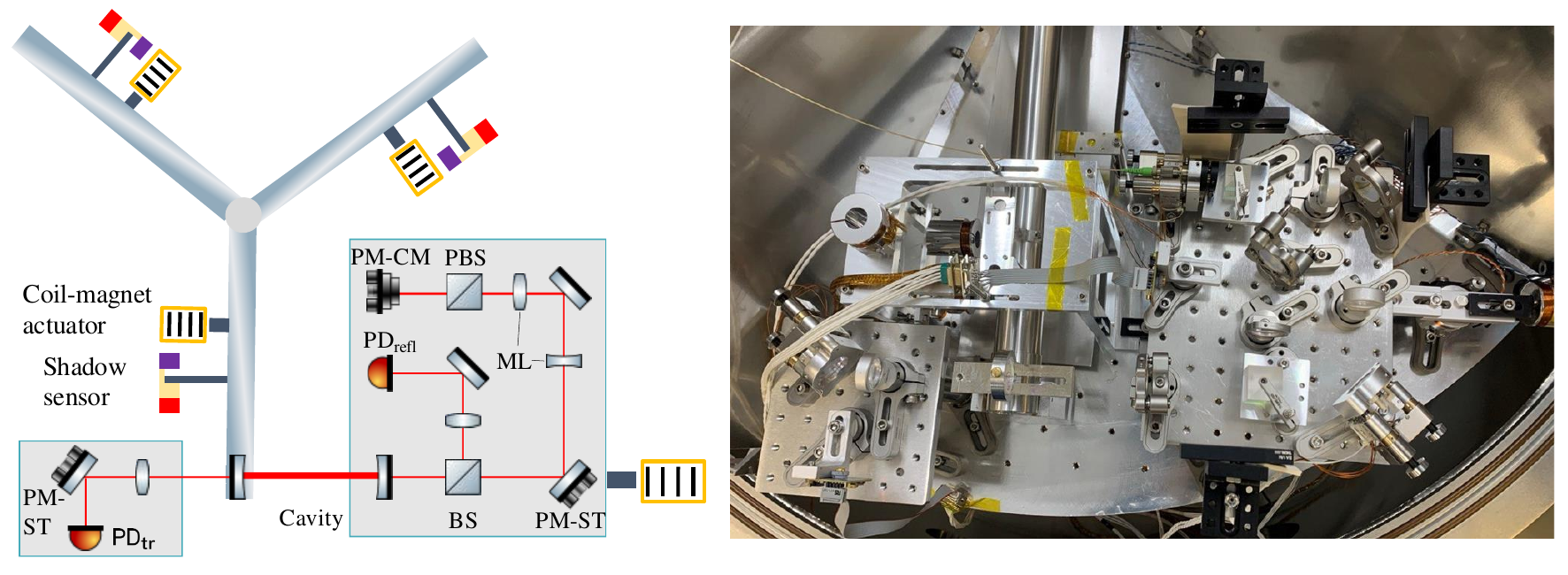}
\caption{\label{fig:epsart} Schematic of the experimental setup on the left; photo of the setup on the right. PM: picomoter, CM: collimator, PBS: polarising beamsplitter, ML: mode-matching lens, ST: steering mirror, BS: beamsplitter, PD: photodetector.}
\end{figure*}

Figure~\ref{fig:epsart} illustrates the in-vacuum setup. The cavity output coupler is fixed on one of the end masses of the seismometer arms. Taking into account the arm length of 0.6\,m, the small rotational differences can be approximated as linear changes in cavity length. The cavity length is \SI{10}{cm}, with an input coupler transmissivity of \SI{8}{ppm} and the round-trip loss is \SI{1}{ppm}. 
%The finesse of the cavity is about \SI{3.5e5}{}. 
This high-finesse design is intended to enhance quantum radiation pressure to excite the SN signal. With a resonant power of about $P_{\rm cav} = 80$\,W in the cavity, the power spectrum of the quantum back action force on the torsion pendulum is \SI{2e-28}{\N^2\per\Hz}.

The cavity input coupler and the laser injection components are set on a board, which is suspended by rubber ribbons, as shown on the right-hand side of Figure~\ref{fig:epsart}.  Two coil-magnet actuators are set in the horizontal and vertical direction on the far right side of the board. The transmission light detection optical components are set on the left board, which rests on three aluminum pillars, supported by rubber pads.

The cavity is powered by a 1550\,nm RIO ORION laser that is located outside the vacuum. The laser beam is delivered into the vacuum through an optical fiber feedthrough, which then directs it into a collimator. A variable optical attenuator is connected between the collimator and the laser and is used for laser intensity stabilization.

We put a polariser in front of the collimator to ensure all the light injected is P-polarized. We used the non-polarized beam splitter to lead the cavity-reflected beam into the photodiode at the reflection port.
We chose not to use a combination of a polarizing beam splitter and a quarter-wave plate \(\lambda/4\) to direct the reflected backward beam to the photodiode because we observed birefringence effects in our Ta$_2$O$_5$ / SiO$_2$ coatings due to the low linewidth of our cavity ($\sim 10^{-6}$ of the cavity free spectral range). The birefringence leads to the non-degeneracy of the two eigenpolarisation states in the cavity. 

For the in-vacuum alignment, we equipped three picomotors and attached them to the laser collimator, one of the steering mirrors in the laser injection part, and the steering mirror before the photodiode in the cavity transmission port. By tuning the pitch and yaw of the injected beam, we managed to keep locking the cavity on a fundamental mode for more than one month.

The eddy currents \cite{plissi2004investigation} induced in the vertical coil frame initially limited the $Q$-factor for the torsion mode to about 1000. After removing the vertical coil, factors such as residual gas, fiber surface, fiber thermoelastic, and the eddy current damping effects from the horizontal coil determined the $Q$-factor. We estimated a $Q$-factor of approximately \SI{5e4}{} for the torsion mode, details can be seen in Appendix \ref{appendix:Q-factor}.

\subsection{Sensing and control system}
We used six servos and one feed-forward for cavity locking to maintain stability and minimize noise: 2 for laser frequency locking, 2 for seismic noise suppression, 1 servo and 1 feed-forward for classical radiation pressure noise suppression, and 1 for intensity stabilization. The control architecture is shown in Figure~\ref{fig:epsart2} in Appendix \ref{appendix:ctrl}. 

The seismic noise induces large mirror motion and strong feedback control is needed to maintain the laser on resonance with the cavity. Thus, we implement a hierarchical locking scheme. The laser frequency is stabilized to follow the resonance of the in-vacuum cavity. This was done using the Pound-Drever-Hall\,\cite{Black_PoundDreverHall_2001} technique, with the signal sourced from the photodetector in the reflection port of the cavity. We inject RF sidebands by modulating the laser frequency at \SI{4}{\MHz} using the laser’s fast current control channel. The error signal is fed back via a dual-channel control scheme to the two supply current channels with a servo bandwidth of \SI{300}{\kHz}. This high bandwidth is necessary for suppressing the relatively high laser frequency noise and catching the narrow error signal due to the high finesse cavity. This laser frequency locking scheme is detailed in Ref.\,\cite{smetana2023reaching, smetana2024high}. However, in this setup, the frequency modulation range is insufficient to maintain the lock indefinitely due to high seismic motion at low frequencies. 

The seismic noise is suppressed in two parts. In the frequency range band from \SI{250}{\mHz} to \SI{10}{\Hz}, the platform of 6D was actively stabilized~\cite{ubhi2022active, prokhorov2023design}. We utilized a shadow sensor with a linear range of \SI{0.7}{\mm} and resolution of \SI{1}{\nm}~\cite{carbone2012sensors} and measured the position between the mass and the platform. This sensor includes a light-emitting diode and a photodiode affixed to the platform, along with a flag attached to the mass. Part of the diode light is blocked by the flag and its absolute position is determined by the photodiode signal.

We suppressed the seismic noise with a bandwidth of \SI{30}{\Hz} by directly modulating the cavity length. As shown in the right panel of Figure~\ref{fig:epsart}, an arm is extended on the right of the board. For both horizontal and vertical directions on the arm, we mounted two \SI{31.4}{\cm^3} Neodymium magnets with a separation of \SI{5}{\mm}. The vertical coil-magnet actuator can provide extra tuning of the pitch of the injected laser beam by pushing the right board in pitch. The horizontal coil-magnet actuates on the board in the direction of the cavity length. When the cavity started to lock with the laser frequency servo, the horizontal coil-magnet actuator was engaged.

We also added a bias servo to suppress the DC offset and low-frequency drifts. This is achieved by modulating the error signal of the cavity length with an \SI{800}{\Hz} sine wave, demodulating the transmitted power, and adding it to the error signal, which is then sent to the laser frequency servo. With the bias servo, we maintain a round-trip detuning phase in the cavity of only \SI{8e-9}{} degree, which is required to suppress optical spring. The resonance frequency shift induced by the optical spring effect is only \SI{5.9}{}\,nHz, which was not resolvable during out observing run that lasted for 3 months. More details are in Appendix \ref{appendix:opticalspring}.

Our hierarchical control stabilises ground vibrations at low frequencies and laser phase noise at higher frequencies. However, upon locking the cavity, a significant radiation pressure force of $2P_{\rm cav}/c=\SI{0.53}{\micro N}$ emerges. If uncompensated, the torsion angle of the suspended mass changes by 233\,mrad and misaligns the cavity. Therefore, we used a shadow sensor to monitor the mass's position, engaged the catching servo to control its movement, and used the coil actuator to push the mass back into position. The catching servo has a high gain from DC, with a unity gain at \SI{2.8}{\mHz}. Apart from the catching servo, we implemented a feed-forward to continuously compensate for the radiation pressure. This is done by sending a control signal proportional to the transmitted power to the coil. Together with the feed-forward and the catching servo, the cavity alignment is preserved during the lock acquisition process.

\subsection{Data analysis}

\begin{figure*}
\includegraphics[height=6.3cm]{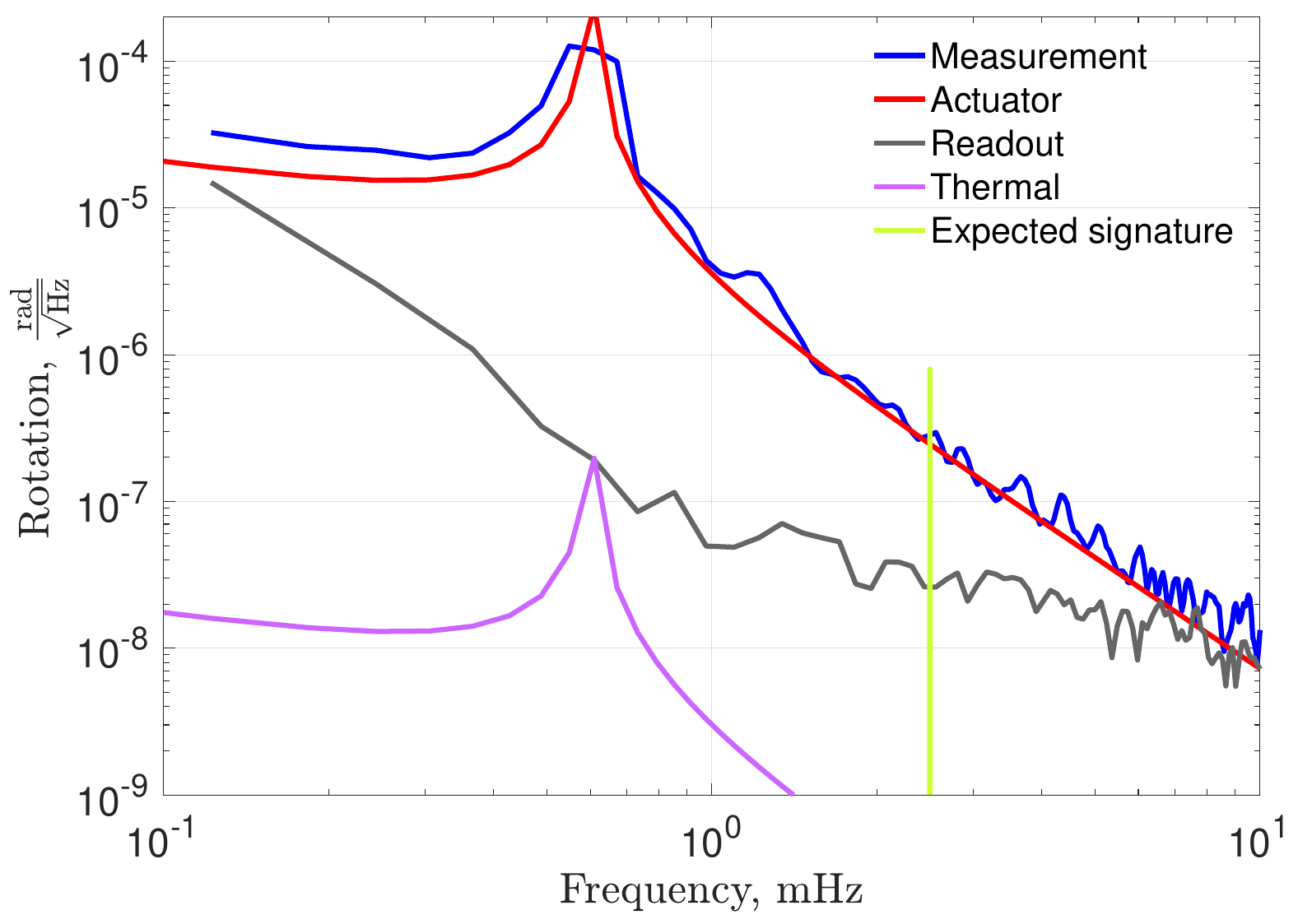} \hspace{3mm}
\includegraphics[height=6.3cm]{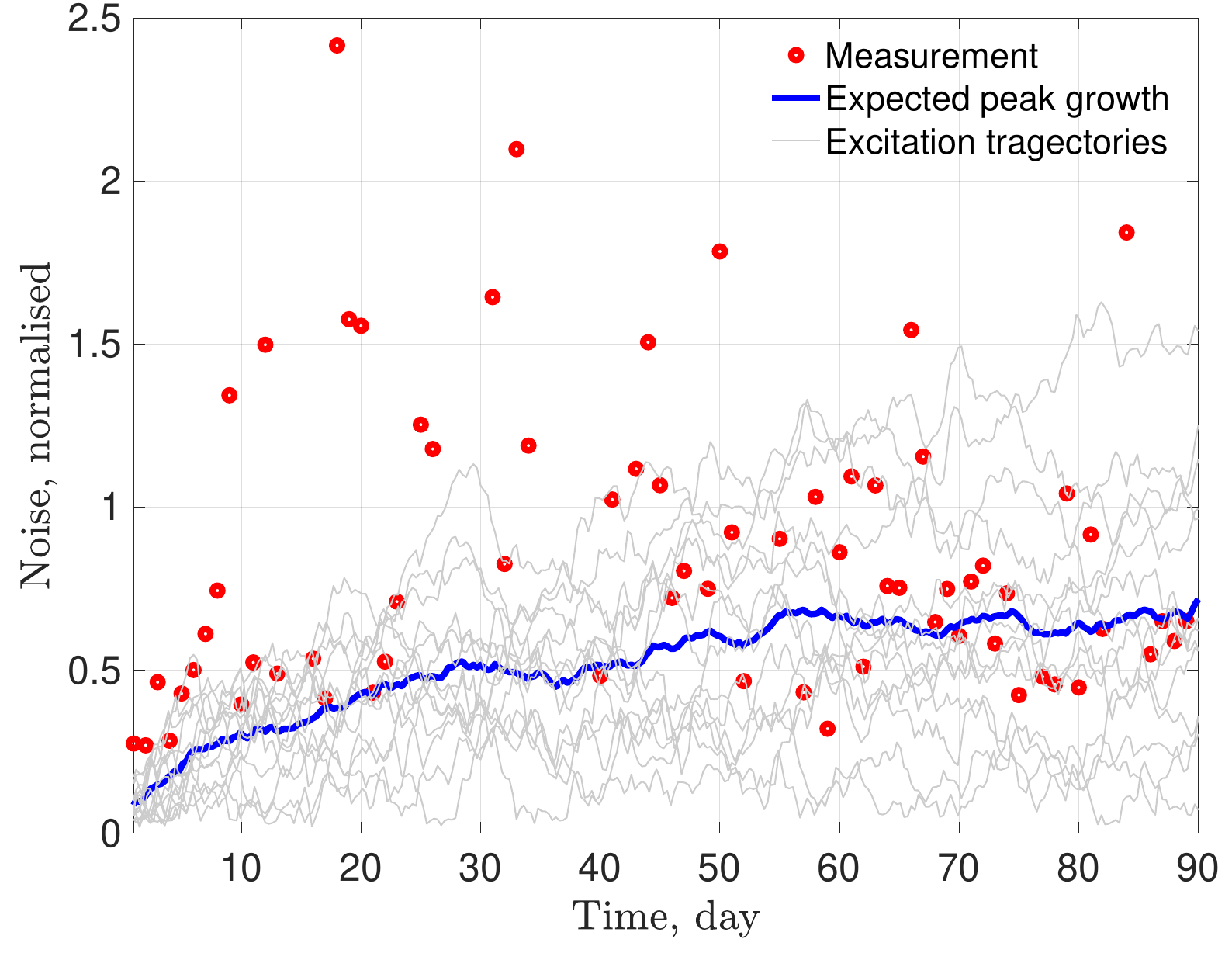}% 

\caption{Left: Torsion motion and sensitivity analysis. The torsion motion is measured by the shadow sensor. The noise is primarily limited by the classical actuator noise below \SI{0.3}{\Hz} and by the readout noise at higher frequencies. The thermal noise of the suspension is significantly lower than our sensitivity.
Right: Result of the data analysis. Red dots show the amplitude spectrum density of the measured data at 2.5\,mHz (where the SN signature is expected under the assumption of pre-selection prescription %casual conditional collapse 
and quadratic SN potential). The spectrum is normalised relative to the value of $\SI{0.3}{\micro rad/\sqrthz}$. The expected peak growth is the average of the simulated excitations of the SN peak over multiple trajectories (shown in gray). Since some of the red dots fall below the blue curve, our sensor has demonstrated the potential to detect the SN signature at 2.5\,mHz.}
\label{Fig:data_analysis} 
\end{figure*}

Figure~\ref{Fig:data_analysis} (left) illustrates the motion spectra of RZ along with the corresponding noise budget. The shadow sensor measures the torsional motion. The noise is primarily limited by the actuator noise and classical radiation pressure noise below 10\,mHz and by readout noise at higher frequencies. The 
thermal noise is not the main contributor.
The cavity control signal is much lower than the shadow sensor signal; however, at the frequency of interest, the system remains limited by actuator force noise around \SI{1}{mHz}. More details can be found in Appendix~\ref{appendix:SN_RZ}. We achieved the sensitivity of $\sqrt{S_{\rm noise}} = \SI{0.3}{\micro rad/\sqrthz}$ at the Schrödinger-Newton frequency of 2.5\,mHz. In the steady state regime considered in~\cite{Helou2017}, the power spectral density of the quantum back action force is given by the equation
\begin{equation}
    S_{FF} = \frac{8 \hbar \omega_0 P_{\rm cav} B}{c^2},
\end{equation}
where $\hbar$ is the reduced Planck constant, $\omega_0$ represents the angular frequency of the laser light, and $B$ is the power build-up in the cavity. Under an assumption of the quadratic SN potential~\cite{yang2013macroscopic}, the power spectral density of the oscillator motion at $\omega_q$ is given by the equation
\begin{equation}
    S_{\theta\theta} = \frac{ Q^2}{I_{\rm RZ}^2 \omega_q^4}L^2S_{FF} ,
\end{equation}
where $I_{\rm RZ}$ is the moment of inertia of the torsion pendulum, $L$ is the length of the arm of the torsion pendulum, and $Q$ is the $Q$-factor. However, $S_{\theta\theta}$ can only be observed if the measurement time is much longer than the inverse of the resonance bandwidth. This time constant is more than 300\,days for our oscillator with a high mechanical quality factor.

While it is possible to observe for more than 300\,days to resolve the peak, slow drifts of the mechanical eigen mode make it impractical to split the data into long segments. The drifts are caused by the parasitic magnetic rigidity of our coil-magnet actuators~\cite{prokhorov2023design}. The rigidity is caused by the small but non-zero derivative of the magnetic force on the angle of the torsion balance. We have minimized the magnetic rigidity by positioning the operating point of the torsion balance such that DC force of the magnets is zero. The approach made the eigenmode stable on the time scale of 1\,day such that we cannot resolve the drifts within the time period. For this reason, we have split our data into segments of $T_{\rm bw}=32768$\,s for data analysis.

In the steady state regime, the power spectral density of the rotational motion is given by the equation
\begin{equation}
    S_{\rm meas} = \frac{\omega_q T_{\rm bw}}{2 \pi Q}S_{\theta\theta}
\end{equation}
and we compute the signal-to-noise ratio of a single window according to the equation
\begin{equation}
    {\rm SNR} = \sqrt{\frac{S_{\rm meas}}{S_{\rm noise}}} = \sqrt{\frac{4}{\pi}\frac{\hbar \omega_0 P_{\rm cav} B T_{\rm bw} Q R^2}{c^2 I_{\rm RZ}^2 \omega_q^3 S_{\rm noise}}} \approx 1
\end{equation}
which shows the potential of our experimental setup to search for the signatures of semiclassical gravity. Figure~\ref{Fig:data_analysis} (left) shows the SN expected signal after the signature is fully excited and under the assumption of the pre-selection prescription %causal conditional collapse 
and of the quadratic SN potential.

However, our torsion balance reaches its steady state only after its ring up (or equally rinddown time) of 300\,days. Here we present the first limitation of measuring features of the semiclassical gravity with high-Q torsion pendulums. Though SNR scales as $\sqrt{Q}$, the ring-up time of Schrödinger-Newton peak is so long that we could not reach the steady state during our observing run of 3 months. We ran simulations of the excitation of the Schrödinger-Newton peak with a quantum back action noise from its quantum thermal state on day 1 of the observing run and compared it with our measured sensitivity. The result is presented in Fig.~\ref{Fig:data_analysis} (right). Due to day-time disturbances on the Birmingham campus, only night-time data was analyzed. However, the cavity was continuously locked during the observing run to keep the excitation of the Schrödinger-Newton peak running.

Our result shows the potential of the setup to measure the Schrödinger-Newton peak even in the transient regime. However, we further discovered that the large uncertainty of the RZ caused by quantum radiation pressure noise broke the quadratic SN potential condition. Using current experimental parameters, the SN signal in the nonquadratic regime and two upgrades are discussed below.
% (Yubao told me this is a bit improper) However, we discovered that the motion of the inertial mass in other rotational and translational degrees of freedom significantly reduces the peak as discussed below.

\section{Future upgrades}
\subsection{Limitations of the current setup}
In the current setup, the uncertainty of the macroscopic RZ rotation excited by quantum radiation pressure can be estimated as $\Delta\Theta= \SI{1.374e-8}{rad}$, as elaborated in Appendix \ref{appendix:macroscopic motion}. With the implementation of the catching servo, this uncertainty can be mitigated to $\Delta\Theta=\SI{5.69e-9}{rad}$ , a value constrained by sensor noise. Considering the atoms situated at the edge of the mass with arm length $L=\SI{0.6}{\m}$, the macroscopic uncertainty of the atom reaches $\Delta X=\SI{3.41e-9}{m}$, which breaks the condition of the quadratic form of SN potential in Eq.\,\ref{eq:Hamiltonian_self_gravity}, $\Delta X\ll\Delta x_{\rm int}$, where $\Delta x_{\rm int}=1.04\times10^{-11}\,{\rm m}$ is the internal displacement uncertainty of the aluminum atom. Consequently, there arises a necessity for identifying a novel signature within this nonquadrature domain. The interaction Hamiltonian governing this non-quadratic regime is:
\begin{equation}
\hat{H}_{\rm SN}(\hat{r}_i)=-\sum_{i,j}Gm_im_j\int\frac{\rho_i(r_i)}{|\hat{r}_i-r_j|}dr_i,
\end{equation}
where $\hat{r}_i=\{\hat{x}_i,\hat{y}_i,\hat{z}_i\}$ is the position operator of the $i_{\rm th}$ atom and $\rho_i(r_i)$ symbolizes the mass distribution attributed to the $i_{\rm th}$ atom, postulated to adhere to a Gaussian profile:
\begin{equation}
\rho_i(x_i,y_i,z_i)=\frac{{\rm exp}\left[-\frac{(x_i-\bar{x}_i)^2}{2\sigma_x^2}-\frac{(y_i-\bar{y}_i)^2}{2\sigma_y^2}-\frac{(z_i-\bar{z}_i)^2}{2\sigma_z^2}\right]}{\sqrt{(2\pi)^3}\sigma_x\sigma_y\sigma_z}.
\end{equation}
where $\sigma_{x/y/z}$ signifies the variance in the positional coordinates of the atom, and $\bar{r}_i$ represents the equilibrium position of the $i$-th atom. A comprehensive analysis of this non-quadratic regime is presented in Appendix \ref{appendix:non-quadrature}. This section merely highlights its core aspects:

(a) The general three-dimensional model simplifies to a one-dimensional effective model, when we assume that the quantum uncertainty of the macroscopic motion arises only from the quantum radiation pressure noise exert soley on the RZ motion. When the mutual self-gravity effect between atoms is negligible, and we ignore the quantum uncertainty of macroscopic motion along other directions to reduce our model to be a one-dimensional problem, the SN Hamiltonian can be empirically written in a parametric form:
\begin{equation}\label{eq:one_dimensional_hamitonian}
H_{\text{SN}}(\Theta) = -\frac{AGMm}{\tilde{r}\sigma_\Theta} \exp\left(-\frac{\Theta^2}{2b_1\sigma^2_\Theta}\right),
\end{equation}
where $\Theta$ represents the macroscopic RZ rotation, $\sigma_\Theta$ is quantum uncertainty of $\Theta$,  $M$ and $m$ are the mass of the pendulum and the atom. The $A$, $b_1$ and the geometric factor $\tilde{r}$ are the empirical coefficients obtained by fitting the simulation results\,(see Appendix \ref{appendix:non-quadrature}). 

(b) The magnitude of the SN signal induced by the self-gravity of each atom decreases as the quantum uncertainty in the macroscopic position increases because $H_{\rm SN}\propto 1/\sigma_\Theta$.

(c) When the fluctuations in macroscopic position increase to the scale of, or exceed, the lattice constant $a$, $\Delta X\geq a$, the mutual gravity between atoms becomes significantly important and severely weakens the SN signal, which is detailed in Appendix \ref{appendix:mutual gravity}.

\begin{figure}
\includegraphics[width=0.48\textwidth]{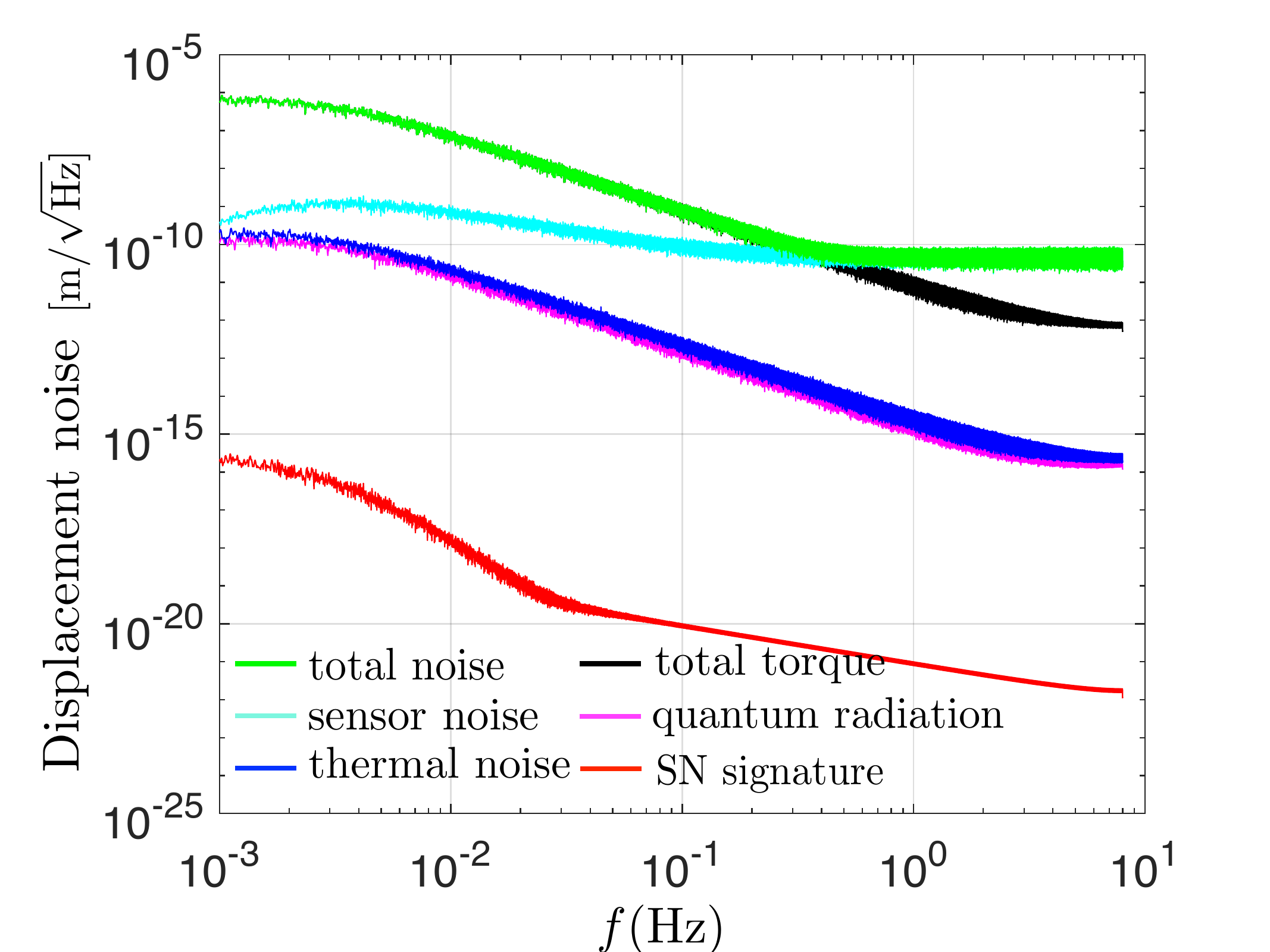}
\includegraphics[width=0.48\textwidth]{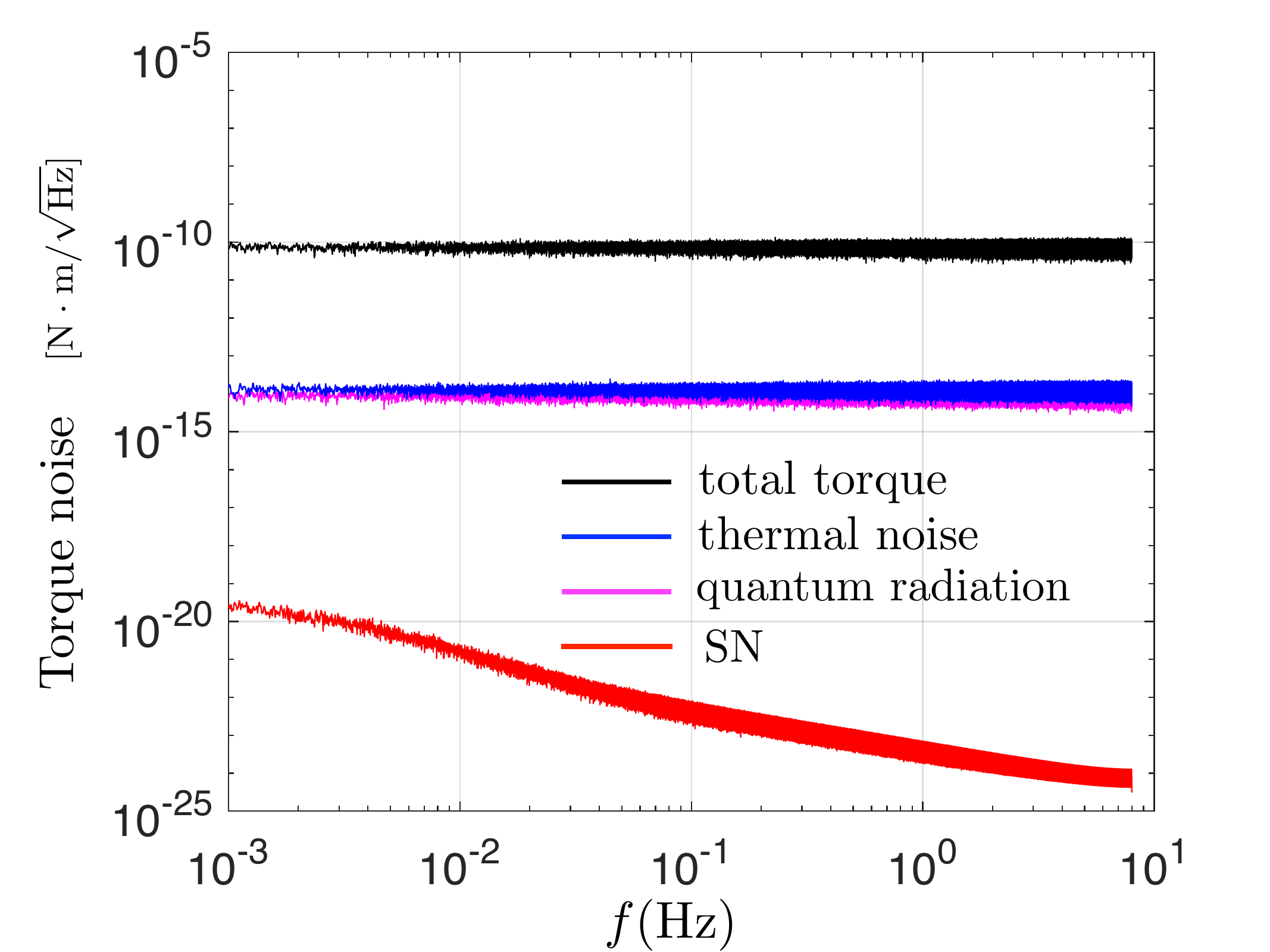}
\caption{\label{fig:spectrum_upper_bound} Simulated displacement noise spectrum (upper panel) and torque noise spectrum (lower panel) for the current experiment. The noise spectrum of SN signal, depicted by the red line, is approximately ten orders of magnitude lower than the total torque noise dipicted by the black line, making the detection of SN gravity challenging within the present experimental setup. }
\end{figure}

Using our numerical model presented in Appendix\,\ref{appendix:non-quadrature}, the parameters $A,b_1,\tilde r$ in the above interaction Hamiltonian can be fitted as $A=3.298, b_1=1.62, \tilde{r}=0.176$. The effect of nonquadratic SN gravity on the measured noise spectrum is simulated and shown in Fig.~\ref{fig:spectrum_upper_bound}.  Note that this spectrum does not have the peak structure as initially anticipated. This is due to (1) the sophisticated servo system that is used to stabilize our system and (2) the nonlinearity of SN dynamics. In standard quantum mechanics, the linearity of the quantum dynamics and the servo loop guarantees that the feedback effect contributes as a multiplication factor to the spectrum without control loops, therefore the feedback effect can be safely removed once we have good modeling of the feedback loop. However, this property is not retained in SN theory due to its nonlinearity, meaning we can only acquire the spectrum that encompasses the feedback effect, which is detailed in Appendix \ref{sec:feedback_role}. Since the Proportional-Integral-Derivative (PID) feedback used in our experimental device would shift the resonant frequency and broaden the linewidth,  the initially anticipated peak structure will not show up in Fig.~\ref{fig:spectrum_upper_bound}.

In the non-quadratic regime, we calculate the maximum bound of the SN signal in the 6D setup based on the current parameters.  From the simulation result, the maximum bound of the SN signal is approximately 10 orders of magnitude lower than the \B{total torque noise}, implying that it would take about $3.02\times10^{14}$ years to distinguish this signal. Therefore, several upgrades are proposed to enable the detection of SN gravity in the 6D setup.

\subsection{Proposed upgrades}
Our previous discussions on the limitations of our setup can be summarized in two main points. First, the significant uncertainty in macroscopic motion renders the quadratic semiclassical gravity (SN) potential inadequate for describing our system. Second, the signal within the non-quadratic potential region is too weak to be detected. To address these challenges, this section proposes possible instrumental upgrades aimed at overcoming these obstacles. The first one aims to enhance our system's capability to test the effect due to quadratic SN potential, while the second one is targeted at testing the non-quadratic SN potential effect.

\subsubsection{Quadratic potential region}
To detect the SN signal in the pre-selection prescription, the key improvements and corresponding strategies are summarized as follows.

(1) \textbf{Suppress the uncertainty of test mass motion.}
Currently, the primary limitation on the precision of test mass motion is the sensor noise. To mitigate this, a highly sensitive sensor with an improvement in sensitivity by four orders of magnitude is necessary. For instance, the current shadow sensor could be replaced with a high-precision optical interferometer. Additionally, enhancing the strength of the feedback system can further suppress the motion of the torsion pendulum; for example, increasing the feedback strength tenfold. Furthermore, augmenting the moment of inertia $I_z$ to $1\,{\rm kg \cdot m^2}$ (for example, increasing the arm length to $1\,{\rm m}$ and concentrating the mass at the periphery of the arm) can reduce the macroscopic uncertainty attributed to quantum radiation pressure noise. With these enhancements, the macroscopic uncertainty is anticipated to be reduced to $3.12\times10^{-12}$\,m.  

\begin{figure}
\includegraphics[width=0.48\textwidth]{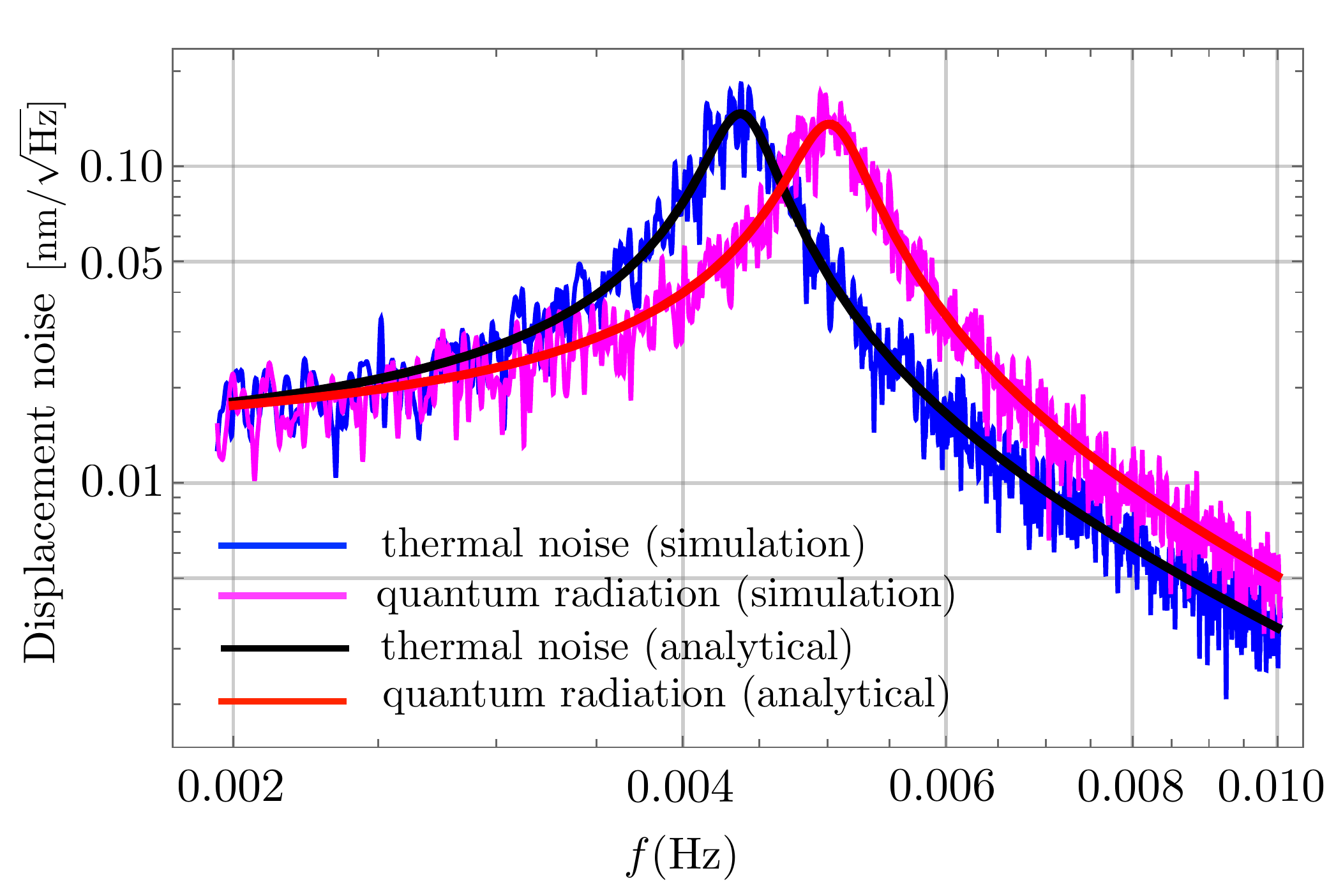}
\caption{\label{fig:SN_spectrum} Simulated noise spectrum in the quadratic regime under the pre-selection prescription. The black and blue lines represent the thermal noise spectrum obtained from analytical solutions and numerical simulations, respectively. The red and magenta lines depict the quantum radiation pressure noise derived from analytical solutions and numerical simulations, respectively. The two-peak structure indicates the signal of SN gravity in the quadratic regime within the pre-selection prescription. Note that the resonant frequency of these two peaks are shifted from $\omega_{\rm m}$ and $\omega_{\rm q}$ by the servo system.}
\end{figure}

(2) \textbf{Optimize the feedback servo to obtain the suitable bandwidth.} As previously discussed, the influence of the feedback system on the measured spectrum cannot be eliminated due to the nonlinear nature of the SN theory. Consequently, its impact must be taken into account when extracting the SN signal from the outgoing optical spectrum. The PID control broadens the bandwidth of the expected peak, and this increased width hinders our ability to differentiate between the mechanical peak and the SN peak. To mitigate the effects of the feedback system, we can introduce a factor $\epsilon$ into the current feedback servo loop. This results in the following transfer function:
\begin{equation}
C(s)=A_0\frac{\epsilon s^2+4\pi\times10^{-3}s+4\pi^2\times10^{-6}}{s(s+2\pi)}.
\end{equation}
Fig.\ref{fig:SN_spectrum} is plotted using the parameter setting $A_0=0.375$ and $\epsilon=0.1$, which demonstrate the avaliability of the revised transfer function.

(3) \textbf{Noise suppression.} In the current setup, there are other types of noise present, such as classical radiation pressure noise, which currently exist at relatively high levels and represent the main sources of classical noise. However, as we improve the sensitivity of our detectors and the servo system, an upgraded feedback mechanism could significantly reduce these classical noises, so that the sensitivity would be limited only by thermal noise, quantum radiation pressure noise, and sensor noise.

\begin{table}[htbp]
    \centering
    \begin{tabular}{|c|c|c|}
    \hline
    Parameters & Symbol & Value \\\hline
    Moment of inertia & $I_{\rm z}$ & \SI{1}{\kg \cdot \m^2}\\
    Mirror bare frequency & $\omega_{\rm m}/2\pi$ & \SI{0.6}{\mHz}\\
    SN frequency & $\omega_{\rm SN}/2\pi$ & \SI{2.53}{\mHz}\\
    Quality factor & $Q$ & \SI{5e5}{}\\
    Optical wavelength & $\lambda$ & \SI{1550}{\nm}\\
    Cavity finesse & $\mathcal{F}$ & \SI{3.5e5}{}\\
    Intra-cavity power & $P_{\rm cav}$ & \SI{80}{\W}\\
    Temperature & $T$ & \SI{300}{\K}\\
    \hline
    \end{tabular}
    \caption{The parameters for the numerical simulation in quadratic SN potential case assuming the pre-selection prescription for quantum measurment. Compared to the case of pure aluminum, the incorporation of brass at the pendulum endpoints results in a shift in the SN frequency. }
    \label{tab:parameter_pre}
\end{table}

With the proposed improvements, we numerically simulate the quadratic SN potential signal under the pre-selection quantum measurement prescription, and the parameters are shown in Table~\ref{tab:parameter_pre}. The SN gravity modulation system exhibits distinct responses to classical noise and quantum noise, resulting in two separate peaks. Both the analytical and simulation results are presented in Figure~\ref{fig:SN_spectrum}.

\subsubsection{Non-quadratic region}
The above-proposed upgrade for testing the quadratic SN gravity effect requires reducing sensor noise by four orders of magnitude, which presents a significant challenge. To mitigate this requirement, in the following section, we explore methods to detect SN gravity in the non-quadratic regime. Furthermore, nonlinear interactions, such as those described in Eq.\,\ref{eq:one_dimensional_hamitonian}, are expected to produce multiple peaks in the noise spectrum, which can be a feature for extracting the SN signal. The proposed upgrades and the simulated results are summarized as follows.

(1) \textbf{Noise suppression.} Similar to the case of testing quadratic SN-potential effect, the system sensitivity should be limited by thermal noise, quantum radiation pressure noise, and sensor noise.  Furthermore, since quantum radiation pressure only drives the degree of freedom of RZ, the quantum uncertainty in the other degrees of freedom is negligible. Therefore, the problem reduces to a one-dimensional case, and the Hamiltonian given in Eq.\,\ref{eq:one_dimensional_hamitonian} becomes valid.

\begin{widetext}

\begin{figure}[t]
\includegraphics[width=1.0\textwidth]{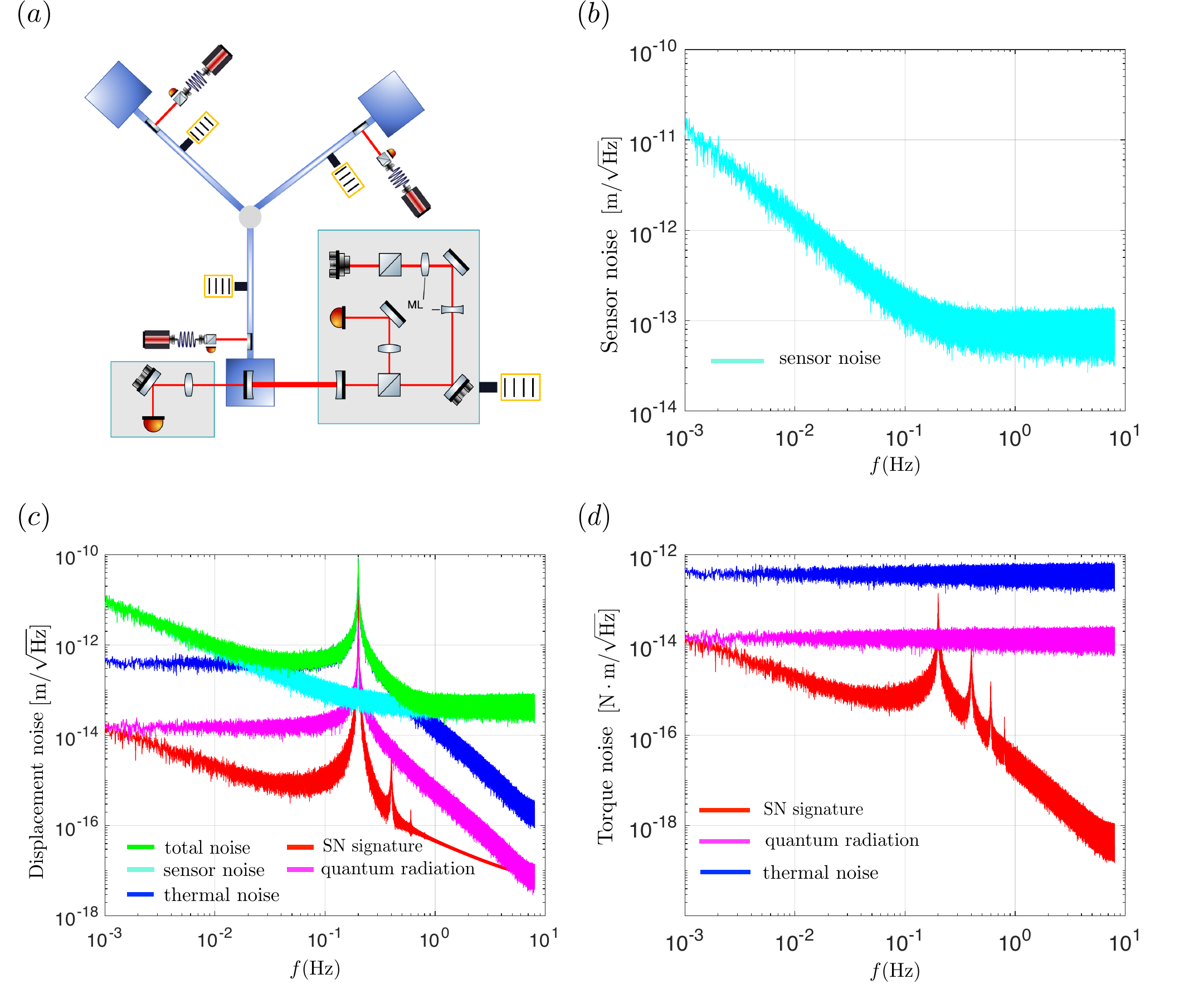}
\caption{\label{fig:SN_spectrum_non}(a) Upgraded experimental setup in the non-quadratic regime, where the mass distribution is concentrated at the end of the arm. (b) sensor noise utilized for the numerical simulation of the SN signal within the non-quadratic regime. (c) Displacement noise spectrum and (d) torque noise spectrum after the upgrade in the non-quadratic regime. Multiple peaks appear in the noise spectrum due to the nonlinear form SN gravity. The SN signature, shown as a red line, is slightly below the torque noise level and can be detected by increasing the measurement time. }
\end{figure}
\end{widetext}

(2) \textbf{Optimize the mass distribution.} The uncertainty in the macroscopic motion must adhere to the condition \( \Delta x_{\rm int} \ll \Delta X \ll a \), where $a$ is the crystal lattice constant of the torsion balance. Here, \(\Delta x_{\rm int} \ll \Delta X \) indicates that the SN potential cannot be quadratic, and \( \Delta X \ll a \) ensures that the motion can not be excessively large, preventing the mutual gravitational attraction between atoms from diminishing the SN effect. In the context of a torsion pendulum, the macroscopic displacement of atoms proximate to the axis of rotation differs significantly from those located further away, with this disparity being directly proportional to the length of the arm. Furthermore, the ratio \( a/\Delta x_{\rm int} \) is constrained to \( \sim 10^2 \), rendering it challenging for all atoms to simultaneously satisfy the condition \( \Delta x_{\rm int} \ll \Delta X \ll a \). To mitigate this challenge, it is imperative to concentrate the mass distribution of the torsion pendulum towards its periphery, as depicted in Fig.~\ref{fig:SN_spectrum_non}.(a).

(3) \textbf{Replace the material by osmium}. The nonquadratic SN potential, as given by Eq.\,\ref{eq:one_dimensional_hamitonian}, is inversely proportional to \( \sigma_\Theta \). Therefore, suppressing the macroscopic displacement uncertainty is beneficial for increasing the SN effect. For the torsion balance shown in Fig.\,\ref{fig:SN_spectrum_non}, smaller rotation uncertainty also means smaller macroscopic displacement uncertainty $\Delta X$ of the test masses situated at the end of the arm.  However, this displacement uncertainty can not be too small and \( \Delta X \gg \Delta x_{\rm int} \) must be satisfied for nonquadratic SN potential. Therefore a smaller atomic internal motion uncertainty $\Delta x_{\rm int}$ allows us to have a lower suppression of $\Delta X$ and hence $\sigma_\Theta$. This can be realised by choosing proper material of the mass such as osmium, of which the atomic internal uncertainty is \( 1 \times 10^{-12} \, \text{m} \), which is ten times smaller than the current one.

(4) \textbf{Increase the resonant frequency of the torsion pendulum.} The feedback system reduces the system's relaxation time while broadening the noise spectrum's bandwidth. When the system's resonant frequency is relatively low, the spacing between multiple peaks excited by SN gravity decreases. If this spacing is smaller than the broadened bandwidth, the multi-peak structure becomes obscured. To reveal this structure, we shift the system's eigenfrequency to \SI{0.2}{\Hz}. To improve the efficacy of the feedback control system at \SI{0.2}{\Hz}, the feedback servo is modified as\,(note that the position of the pole is shifted for stabilization): 
\begin{equation}
C(s) = A_0 \frac{\epsilon s^2 + 4\pi \times 10^{-2}s + 4\pi^2 \times 10^{-4}}{s(s + 20\pi)},
\end{equation}
where $A_0 = 0.376$, $\epsilon = 0.5$, and note that the numerical parameters in the numerator are also adjusted accordingly to obtain the optimal feedback servo function.

\begin{table}[htbp]
    \centering
    \begin{tabular}{|c|c|c|}
    \hline
    Parameters & Symbol & Value \\\hline
    Moment of inertia & $I_{\rm z}$ & \SI{0.36}{\kg \cdot \m^2}\\
    Mirror bare frequency & $\omega_{\rm m}/2\pi$ & \SI{0.2}{\Hz}\\
    Quality factor & $Q$ & \SI{5e4}{}\\
    Arm length & $L$ & \SI{0.6}{\m}\\
    Optical wavelength & $\lambda$ & \SI{1550}{\nm}\\
    Cavity finesse & $\mathcal{F}$ & \SI{3.5e5}{}\\
    Intra-cavity power & $P_{\rm cav}$ & \SI{80}{\W}\\
    Temperature & $T$ & \SI{300}{\K}\\
    \hline
    \end{tabular}
    \caption{The parameters for the numerical simulation in the non-quadratic regime.}
    \label{tab:parameter_nq}
\end{table}

Besides, a detector with sensor noise three orders of magnitude lower than the current one is needed, which is a less stringent requirement than in the quadratic SN-potential case. Compact Michelson interferometers can already\,\cite{smetana2022compact} offer two orders of magnitude better sensitivity at \SI{0.2}{Hz} compared to the shadow sensor we use currently. A feedback system to reduce the macroscopic uncertainty to \SI{1.09e-11}{\m} is also assumed. Considering the aforementioned upgrades, the numerical simulation of the SN signal is shown in Fig.\,\ref{fig:SN_spectrum_non} with the parameters in Table.\,\ref{tab:parameter_nq}. The nonlinear SN interactions excite multiple peaks in the noise spectrum. Although the SN signal is relatively weak and below the thermal noise level, we can improve the signal-to-noise ratio by identifying optimized estimators and increasing the measurement time. It is estimated that with a measurement time of approximately 0.5 years, a signal-to-noise ratio of unity can be achieved.

\section{Conclusion}
We have developed a low-mechanical frequency optomechanical system tailored for the investigation of the semiclassical gravity effect. Our design incorporates a cavity with finesse over \SI{3.5e5}{} to enhance quantum radiation pressure and thus amplify the signal predicted by the SN theory. The test mass, featuring a torsion resonance frequency of \SI{0.6}{\mHz}, is optimized to detect the most prominent SN signature. We estimate that our pendulum achieves a $Q$-factor of at least \SI{5e4}{} to maximize the SNR.

To ensure the stability of the cavity locking and minimize noise, we implemented a sophisticated control architecture. This includes two servos for laser frequency locking, two additional servos for seismic noise suppression, one servo coupled with a feed-forward for classical radiation pressure noise suppression, and another servo for laser intensity stabilization. With this comprehensive control strategy, we achieved a sensing sensitivity of \SI{0.3}{\micro rad/\sqrthz} at \SI{2.5}{mHz}.

We demonstrate the sensitivity capable of observing semiclassical gravity signatures in the quadratic SN potential and analyzed 3 months of data. We have also identified a list of technical challenges with testing SN signatures with torsion balances, including optical springs and hierarchical controls. We also found that our SN signature was much smaller than originally expected because we operated at the non-quadratic regime of the SN theory. We developed a new formalism to analyze non-quadratic potentials and discuss opportunities for further improvements of the instrument.

For both the quadratic and the non-quadratic regimes of SN theory, future enhancements should include: (1) Employing more sensitive sensors to improve detection capabilities. (2) Upgrading feedback servos to reduce the uncertainty in macroscopic displacement. Notably, for the non-quadrature regime, optimizing the mass distribution and the resonance frequency of the torsion pendulum and replacing the material of the test mass with osmium could yield significant benefits. Preliminary estimates suggest that with approximately 0.5 years of measurement time, achieving a signal-to-noise ratio of unity is feasible.

\section{Acknowledgement}
We acknowledge members of the UK Quantum Interferometry collaboration for useful discussions, the support of the Institute for Gravitational Wave Astronomy at the University of Birmingham and STFC Quantum Technology for Fundamental Physics scheme (Grant No. ST/T006609/1 and ST/W006375/1). Y.M and Y.L. thanks Professor Yanbei Chen for useful conversations. D.M. is supported by the 2021 Philip Leverhulme Prize.  Y.M. is supported by the National Key R$\&$D Program of China ``Gravitational Wave Detection"
(Grant No.: 2023YFC2205801), and National Natural Science Foundation of China under Grant No.12474481. H.M. would like to acknowledge the support from the National Key R$\&$D Program of China ``Gravitational Wave Detection" (Grant No.: 2023YFC2205800), and support from Quantum Information Frontier Science Center.

\bibliography{main}% Produces the bibliography via BibTeX

\providecommand{\noopsort}[1]{}\providecommand{\singleletter}[1]{#1}%
\begin{thebibliography}{33}%
\makeatletter
\providecommand \@ifxundefined [1]{%
 \@ifx{#1\undefined}
}%
\providecommand \@ifnum [1]{%
 \ifnum #1\expandafter \@firstoftwo
 \else \expandafter \@secondoftwo
 \fi
}%
\providecommand \@ifx [1]{%
 \ifx #1\expandafter \@firstoftwo
 \else \expandafter \@secondoftwo
 \fi
}%
\providecommand \natexlab [1]{#1}%
\providecommand \enquote  [1]{``#1''}%
\providecommand \bibnamefont  [1]{#1}%
\providecommand \bibfnamefont [1]{#1}%
\providecommand \citenamefont [1]{#1}%
\providecommand \href@noop [0]{\@secondoftwo}%
\providecommand \href [0]{\begingroup \@sanitize@url \@href}%
\providecommand \@href[1]{\@@startlink{#1}\@@href}%
\providecommand \@@href[1]{\endgroup#1\@@endlink}%
\providecommand \@sanitize@url [0]{\catcode `\\12\catcode `\$12\catcode `\&12\catcode `\#12\catcode `\^12\catcode `\_12\catcode `\%12\relax}%
\providecommand \@@startlink[1]{}%
\providecommand \@@endlink[0]{}%
\providecommand \url  [0]{\begingroup\@sanitize@url \@url }%
\providecommand \@url [1]{\endgroup\@href {#1}{\urlprefix }}%
\providecommand \urlprefix  [0]{URL }%
\providecommand \Eprint [0]{\href }%
\providecommand \doibase [0]{https://doi.org/}%
\providecommand \selectlanguage [0]{\@gobble}%
\providecommand \bibinfo  [0]{\@secondoftwo}%
\providecommand \bibfield  [0]{\@secondoftwo}%
\providecommand \translation [1]{[#1]}%
\providecommand \BibitemOpen [0]{}%
\providecommand \bibitemStop [0]{}%
\providecommand \bibitemNoStop [0]{.\EOS\space}%
\providecommand \EOS [0]{\spacefactor3000\relax}%
\providecommand \BibitemShut  [1]{\csname bibitem#1\endcsname}%
\let\auto@bib@innerbib\@empty
%</preamble>
\bibitem [{\citenamefont {{Rosenfeld}}(1963)}]{Rosenfeld1963}%
  \BibitemOpen
  \bibfield  {author} {\bibinfo {author} {\bibfnamefont {L.}~\bibnamefont {{Rosenfeld}}},\ }\bibfield  {title} {\bibinfo {title} {{On quantization of fields}},\ }\href {https://doi.org/10.1016/0029-5582(63)90279-7} {\bibfield  {journal} {\bibinfo  {journal} {Nuclear Physics}\ }\textbf {\bibinfo {volume} {40}},\ \bibinfo {pages} {353} (\bibinfo {year} {1963})}\BibitemShut {NoStop}%
\bibitem [{\citenamefont {{Mueller C}}(1962)}]{Mueller1962}%
  \BibitemOpen
  \bibfield  {author} {\bibinfo {author} {\bibnamefont {{Mueller C}}},\ }\href@noop {} {\emph {\bibinfo {title} {{Les The'ories Relativistes de la Gravitation (Colloques Internationaux CNRS), edited by A Lichnerowicz and M-A Tonnelat}}}}\ (\bibinfo  {publisher} {Paris: CNRS},\ \bibinfo {year} {1962})\BibitemShut {NoStop}%
\bibitem [{\citenamefont {Simon}\ \emph {et~al.}(2001)\citenamefont {Simon}, \citenamefont {Bu\ifmmode~\check{z}\else \v{z}\fi{}ek},\ and\ \citenamefont {Gisin}}]{Simon_2001}%
  \BibitemOpen
  \bibfield  {author} {\bibinfo {author} {\bibfnamefont {C.}~\bibnamefont {Simon}}, \bibinfo {author} {\bibfnamefont {V.}~\bibnamefont {Bu\ifmmode~\check{z}\else \v{z}\fi{}ek}},\ and\ \bibinfo {author} {\bibfnamefont {N.}~\bibnamefont {Gisin}},\ }\bibfield  {title} {\bibinfo {title} {No-signaling condition and quantum dynamics},\ }\href {https://doi.org/10.1103/PhysRevLett.87.170405} {\bibfield  {journal} {\bibinfo  {journal} {Phys. Rev. Lett.}\ }\textbf {\bibinfo {volume} {87}},\ \bibinfo {pages} {170405} (\bibinfo {year} {2001})}\BibitemShut {NoStop}%
\bibitem [{\citenamefont {Mielnik}(2001)}]{Mielnik_20011}%
  \BibitemOpen
  \bibfield  {author} {\bibinfo {author} {\bibfnamefont {B.}~\bibnamefont {Mielnik}},\ }\bibfield  {title} {\bibinfo {title} {Nonlinear quantum mechanics: a conflict with the ptolomean structure?},\ }\href {https://doi.org/https://doi.org/10.1016/S0375-9601(01)00583-7} {\bibfield  {journal} {\bibinfo  {journal} {Physics Letters A}\ }\textbf {\bibinfo {volume} {289}},\ \bibinfo {pages} {1} (\bibinfo {year} {2001})}\BibitemShut {NoStop}%
\bibitem [{\citenamefont {Jayich}\ \emph {et~al.}(2008)\citenamefont {Jayich}, \citenamefont {Sankey}, \citenamefont {Zwickl}, \citenamefont {Yang}, \citenamefont {Thompson}, \citenamefont {Girvin}, \citenamefont {Clerk}, \citenamefont {Marquardt},\ and\ \citenamefont {Harris}}]{Jayich2008}%
  \BibitemOpen
  \bibfield  {author} {\bibinfo {author} {\bibfnamefont {A.~M.}\ \bibnamefont {Jayich}}, \bibinfo {author} {\bibfnamefont {J.~C.}\ \bibnamefont {Sankey}}, \bibinfo {author} {\bibfnamefont {B.~M.}\ \bibnamefont {Zwickl}}, \bibinfo {author} {\bibfnamefont {C.}~\bibnamefont {Yang}}, \bibinfo {author} {\bibfnamefont {J.~D.}\ \bibnamefont {Thompson}}, \bibinfo {author} {\bibfnamefont {S.~M.}\ \bibnamefont {Girvin}}, \bibinfo {author} {\bibfnamefont {A.~A.}\ \bibnamefont {Clerk}}, \bibinfo {author} {\bibfnamefont {F.}~\bibnamefont {Marquardt}},\ and\ \bibinfo {author} {\bibfnamefont {J.~G.~E.}\ \bibnamefont {Harris}},\ }\bibfield  {title} {\bibinfo {title} {Dispersive optomechanics: a membrane inside a cavity},\ }\href {https://doi.org/10.1088/1367-2630/10/9/095008} {\bibfield  {journal} {\bibinfo  {journal} {New Journal of Physics}\ }\textbf {\bibinfo {volume} {10}},\ \bibinfo {pages} {095008} (\bibinfo {year} {2008})}\BibitemShut {NoStop}%
\bibitem [{\citenamefont {Chen}(2013)}]{Chen_2013}%
  \BibitemOpen
  \bibfield  {author} {\bibinfo {author} {\bibfnamefont {Y.}~\bibnamefont {Chen}},\ }\bibfield  {title} {\bibinfo {title} {Macroscopic quantum mechanics: theory and experimental concepts of optomechanics},\ }\href {https://doi.org/10.1088/0953-4075/46/10/104001} {\bibfield  {journal} {\bibinfo  {journal} {Journal of Physics B: Atomic, Molecular and Optical Physics}\ }\textbf {\bibinfo {volume} {46}},\ \bibinfo {pages} {104001} (\bibinfo {year} {2013})}\BibitemShut {NoStop}%
\bibitem [{\citenamefont {{Aspelmeyer}}\ \emph {et~al.}(2014)\citenamefont {{Aspelmeyer}}, \citenamefont {{Kippenberg}},\ and\ \citenamefont {{Marquardt}}}]{Aspelmyer2014}%
  \BibitemOpen
  \bibfield  {author} {\bibinfo {author} {\bibfnamefont {M.}~\bibnamefont {{Aspelmeyer}}}, \bibinfo {author} {\bibfnamefont {T.~J.}\ \bibnamefont {{Kippenberg}}},\ and\ \bibinfo {author} {\bibfnamefont {F.}~\bibnamefont {{Marquardt}}},\ }\bibfield  {title} {\bibinfo {title} {{Cavity optomechanics}},\ }\href {https://doi.org/10.1103/RevModPhys.86.1391} {\bibfield  {journal} {\bibinfo  {journal} {Reviews of Modern Physics}\ }\textbf {\bibinfo {volume} {86}},\ \bibinfo {pages} {1391} (\bibinfo {year} {2014})},\ \Eprint {https://arxiv.org/abs/1303.0733} {arXiv:1303.0733 [cond-mat.mes-hall]} \BibitemShut {NoStop}%
\bibitem [{\citenamefont {Hoang}\ \emph {et~al.}(2016)\citenamefont {Hoang}, \citenamefont {Ma}, \citenamefont {Ahn}, \citenamefont {Bang}, \citenamefont {Robicheaux}, \citenamefont {Yin},\ and\ \citenamefont {Li}}]{Hoang2016}%
  \BibitemOpen
  \bibfield  {author} {\bibinfo {author} {\bibfnamefont {T.~M.}\ \bibnamefont {Hoang}}, \bibinfo {author} {\bibfnamefont {Y.}~\bibnamefont {Ma}}, \bibinfo {author} {\bibfnamefont {J.}~\bibnamefont {Ahn}}, \bibinfo {author} {\bibfnamefont {J.}~\bibnamefont {Bang}}, \bibinfo {author} {\bibfnamefont {F.}~\bibnamefont {Robicheaux}}, \bibinfo {author} {\bibfnamefont {Z.-Q.}\ \bibnamefont {Yin}},\ and\ \bibinfo {author} {\bibfnamefont {T.}~\bibnamefont {Li}},\ }\bibfield  {title} {\bibinfo {title} {Torsional optomechanics of a levitated nonspherical nanoparticle},\ }\href {https://doi.org/10.1103/PhysRevLett.117.123604} {\bibfield  {journal} {\bibinfo  {journal} {Phys. Rev. Lett.}\ }\textbf {\bibinfo {volume} {117}},\ \bibinfo {pages} {123604} (\bibinfo {year} {2016})}\BibitemShut {NoStop}%
\bibitem [{\citenamefont {Rossi}\ \emph {et~al.}(2018)\citenamefont {Rossi}, \citenamefont {Mason}, \citenamefont {Chen}, \citenamefont {Tsaturyan},\ and\ \citenamefont {Schliesser}}]{Rossi2018}%
  \BibitemOpen
  \bibfield  {author} {\bibinfo {author} {\bibfnamefont {M.}~\bibnamefont {Rossi}}, \bibinfo {author} {\bibfnamefont {D.}~\bibnamefont {Mason}}, \bibinfo {author} {\bibfnamefont {J.}~\bibnamefont {Chen}}, \bibinfo {author} {\bibfnamefont {Y.}~\bibnamefont {Tsaturyan}},\ and\ \bibinfo {author} {\bibfnamefont {A.}~\bibnamefont {Schliesser}},\ }\bibfield  {title} {\bibinfo {title} {Measurement-based quantum control of mechanical motion},\ }\href {https://doi.org/10.1038/s41586-018-0643-8} {\bibfield  {journal} {\bibinfo  {journal} {Nature}\ }\textbf {\bibinfo {volume} {563}},\ \bibinfo {pages} {53} (\bibinfo {year} {2018})}\BibitemShut {NoStop}%
\bibitem [{\citenamefont {Mason}\ \emph {et~al.}(2019)\citenamefont {Mason}, \citenamefont {Chen}, \citenamefont {Rossi}, \citenamefont {Tsaturyan},\ and\ \citenamefont {Schliesser}}]{Mason2019}%
  \BibitemOpen
  \bibfield  {author} {\bibinfo {author} {\bibfnamefont {D.}~\bibnamefont {Mason}}, \bibinfo {author} {\bibfnamefont {J.}~\bibnamefont {Chen}}, \bibinfo {author} {\bibfnamefont {M.}~\bibnamefont {Rossi}}, \bibinfo {author} {\bibfnamefont {Y.}~\bibnamefont {Tsaturyan}},\ and\ \bibinfo {author} {\bibfnamefont {A.}~\bibnamefont {Schliesser}},\ }\bibfield  {title} {\bibinfo {title} {Continuous force and displacement measurement below the standard quantum limit},\ }\href {https://doi.org/10.1038/s41567-019-0533-5} {\bibfield  {journal} {\bibinfo  {journal} {Nature Physics}\ }\textbf {\bibinfo {volume} {15}},\ \bibinfo {pages} {745} (\bibinfo {year} {2019})}\BibitemShut {NoStop}%
\bibitem [{\citenamefont {Yu}\ \emph {et~al.}(2020)\citenamefont {Yu}, \citenamefont {McCuller}, \citenamefont {Tse}, \citenamefont {Kijbunchoo}, \citenamefont {Barsotti}, \citenamefont {Mavalvala},\ and\ \citenamefont {members of~the LIGO Scientific~Collaboration}}]{Yu2020}%
  \BibitemOpen
  \bibfield  {author} {\bibinfo {author} {\bibfnamefont {H.}~\bibnamefont {Yu}}, \bibinfo {author} {\bibfnamefont {L.}~\bibnamefont {McCuller}}, \bibinfo {author} {\bibfnamefont {M.}~\bibnamefont {Tse}}, \bibinfo {author} {\bibfnamefont {N.}~\bibnamefont {Kijbunchoo}}, \bibinfo {author} {\bibfnamefont {L.}~\bibnamefont {Barsotti}}, \bibinfo {author} {\bibfnamefont {N.}~\bibnamefont {Mavalvala}},\ and\ \bibinfo {author} {\bibnamefont {members of~the LIGO Scientific~Collaboration}},\ }\bibfield  {title} {\bibinfo {title} {Quantum correlations between light and the kilogram-mass mirrors of ligo},\ }\href {https://doi.org/10.1038/s41586-020-2420-8} {\bibfield  {journal} {\bibinfo  {journal} {Nature}\ }\textbf {\bibinfo {volume} {583}},\ \bibinfo {pages} {43} (\bibinfo {year} {2020})}\BibitemShut {NoStop}%
\bibitem [{\citenamefont {Aggarwal}\ \emph {et~al.}(2020)\citenamefont {Aggarwal}, \citenamefont {Cullen}, \citenamefont {Cripe}, \citenamefont {Cole}, \citenamefont {Lanza}, \citenamefont {Libson}, \citenamefont {Follman}, \citenamefont {Heu}, \citenamefont {Corbitt},\ and\ \citenamefont {Mavalvala}}]{Aggarwal2020}%
  \BibitemOpen
  \bibfield  {author} {\bibinfo {author} {\bibfnamefont {N.}~\bibnamefont {Aggarwal}}, \bibinfo {author} {\bibfnamefont {T.~J.}\ \bibnamefont {Cullen}}, \bibinfo {author} {\bibfnamefont {J.}~\bibnamefont {Cripe}}, \bibinfo {author} {\bibfnamefont {G.~D.}\ \bibnamefont {Cole}}, \bibinfo {author} {\bibfnamefont {R.}~\bibnamefont {Lanza}}, \bibinfo {author} {\bibfnamefont {A.}~\bibnamefont {Libson}}, \bibinfo {author} {\bibfnamefont {D.}~\bibnamefont {Follman}}, \bibinfo {author} {\bibfnamefont {P.}~\bibnamefont {Heu}}, \bibinfo {author} {\bibfnamefont {T.}~\bibnamefont {Corbitt}},\ and\ \bibinfo {author} {\bibfnamefont {N.}~\bibnamefont {Mavalvala}},\ }\bibfield  {title} {\bibinfo {title} {Room-temperature optomechanical squeezing},\ }\href {https://doi.org/10.1038/s41567-020-0877-x} {\bibfield  {journal} {\bibinfo  {journal} {Nature Physics}\ }\textbf {\bibinfo {volume} {16}},\ \bibinfo {pages} {784} (\bibinfo {year} {2020})}\BibitemShut {NoStop}%
\bibitem [{\citenamefont {Cata\~no Lopez}\ \emph {et~al.}(2020)\citenamefont {Cata\~no Lopez}, \citenamefont {Santiago-Condori}, \citenamefont {Edamatsu},\ and\ \citenamefont {Matsumoto}}]{Cata2020}%
  \BibitemOpen
  \bibfield  {author} {\bibinfo {author} {\bibfnamefont {S.~B.}\ \bibnamefont {Cata\~no Lopez}}, \bibinfo {author} {\bibfnamefont {J.~G.}\ \bibnamefont {Santiago-Condori}}, \bibinfo {author} {\bibfnamefont {K.}~\bibnamefont {Edamatsu}},\ and\ \bibinfo {author} {\bibfnamefont {N.}~\bibnamefont {Matsumoto}},\ }\bibfield  {title} {\bibinfo {title} {High-$q$ milligram-scale monolithic pendulum for quantum-limited gravity measurements},\ }\href {https://doi.org/10.1103/PhysRevLett.124.221102} {\bibfield  {journal} {\bibinfo  {journal} {Phys. Rev. Lett.}\ }\textbf {\bibinfo {volume} {124}},\ \bibinfo {pages} {221102} (\bibinfo {year} {2020})}\BibitemShut {NoStop}%
\bibitem [{\citenamefont {Helou}\ \emph {et~al.}(2017)\citenamefont {Helou}, \citenamefont {Luo}, \citenamefont {Yeh}, \citenamefont {Shao}, \citenamefont {Slagmolen}, \citenamefont {McClelland},\ and\ \citenamefont {Chen}}]{Helou2017}%
  \BibitemOpen
  \bibfield  {author} {\bibinfo {author} {\bibfnamefont {B.}~\bibnamefont {Helou}}, \bibinfo {author} {\bibfnamefont {J.}~\bibnamefont {Luo}}, \bibinfo {author} {\bibfnamefont {H.-C.}\ \bibnamefont {Yeh}}, \bibinfo {author} {\bibfnamefont {C.-g.}\ \bibnamefont {Shao}}, \bibinfo {author} {\bibfnamefont {B.~J.~J.}\ \bibnamefont {Slagmolen}}, \bibinfo {author} {\bibfnamefont {D.~E.}\ \bibnamefont {McClelland}},\ and\ \bibinfo {author} {\bibfnamefont {Y.}~\bibnamefont {Chen}},\ }\bibfield  {title} {\bibinfo {title} {Measurable signatures of quantum mechanics in a classical spacetime},\ }\href {https://doi.org/10.1103/PhysRevD.96.044008} {\bibfield  {journal} {\bibinfo  {journal} {Phys. Rev. D}\ }\textbf {\bibinfo {volume} {96}},\ \bibinfo {pages} {044008} (\bibinfo {year} {2017})}\BibitemShut {NoStop}%
\bibitem [{\citenamefont {Yang}\ \emph {et~al.}(2013)\citenamefont {Yang}, \citenamefont {Miao}, \citenamefont {Lee}, \citenamefont {Helou},\ and\ \citenamefont {Chen}}]{yang2013macroscopic}%
  \BibitemOpen
  \bibfield  {author} {\bibinfo {author} {\bibfnamefont {H.}~\bibnamefont {Yang}}, \bibinfo {author} {\bibfnamefont {H.}~\bibnamefont {Miao}}, \bibinfo {author} {\bibfnamefont {D.-S.}\ \bibnamefont {Lee}}, \bibinfo {author} {\bibfnamefont {B.}~\bibnamefont {Helou}},\ and\ \bibinfo {author} {\bibfnamefont {Y.}~\bibnamefont {Chen}},\ }\bibfield  {title} {\bibinfo {title} {Macroscopic quantum mechanics in a classical spacetime},\ }\href@noop {} {\bibfield  {journal} {\bibinfo  {journal} {Physical review letters}\ }\textbf {\bibinfo {volume} {110}},\ \bibinfo {pages} {170401} (\bibinfo {year} {2013})}\BibitemShut {NoStop}%
\bibitem [{\citenamefont {Liu}\ \emph {et~al.}(2023)\citenamefont {Liu}, \citenamefont {Miao}, \citenamefont {Chen},\ and\ \citenamefont {Ma}}]{Liu2023}%
  \BibitemOpen
  \bibfield  {author} {\bibinfo {author} {\bibfnamefont {Y.}~\bibnamefont {Liu}}, \bibinfo {author} {\bibfnamefont {H.}~\bibnamefont {Miao}}, \bibinfo {author} {\bibfnamefont {Y.}~\bibnamefont {Chen}},\ and\ \bibinfo {author} {\bibfnamefont {Y.}~\bibnamefont {Ma}},\ }\bibfield  {title} {\bibinfo {title} {Semiclassical gravity phenomenology under the causal-conditional quantum measurement prescription},\ }\href {https://doi.org/10.1103/PhysRevD.107.024004} {\bibfield  {journal} {\bibinfo  {journal} {Phys. Rev. D}\ }\textbf {\bibinfo {volume} {107}},\ \bibinfo {pages} {024004} (\bibinfo {year} {2023})}\BibitemShut {NoStop}%
\bibitem [{\citenamefont {Mow-Lowry}\ and\ \citenamefont {Martynov}(2019)}]{mow20196d}%
  \BibitemOpen
  \bibfield  {author} {\bibinfo {author} {\bibfnamefont {C.~M.}\ \bibnamefont {Mow-Lowry}}\ and\ \bibinfo {author} {\bibfnamefont {D.}~\bibnamefont {Martynov}},\ }\bibfield  {title} {\bibinfo {title} {A 6d interferometric inertial isolation system},\ }\href@noop {} {\bibfield  {journal} {\bibinfo  {journal} {Classical and Quantum Gravity}\ }\textbf {\bibinfo {volume} {36}},\ \bibinfo {pages} {245006} (\bibinfo {year} {2019})}\BibitemShut {NoStop}%
\bibitem [{\citenamefont {Ubhi}\ \emph {et~al.}(2022)\citenamefont {Ubhi}, \citenamefont {Prokhorov}, \citenamefont {Cooper}, \citenamefont {Fronzo}, \citenamefont {Bryant}, \citenamefont {Hoyland}, \citenamefont {Mitchell}, \citenamefont {Van~Dongen}, \citenamefont {Mow-Lowry}, \citenamefont {Cumming} \emph {et~al.}}]{ubhi2022active}%
  \BibitemOpen
  \bibfield  {author} {\bibinfo {author} {\bibfnamefont {A.~S.}\ \bibnamefont {Ubhi}}, \bibinfo {author} {\bibfnamefont {L.}~\bibnamefont {Prokhorov}}, \bibinfo {author} {\bibfnamefont {S.}~\bibnamefont {Cooper}}, \bibinfo {author} {\bibfnamefont {C.~D.}\ \bibnamefont {Fronzo}}, \bibinfo {author} {\bibfnamefont {J.}~\bibnamefont {Bryant}}, \bibinfo {author} {\bibfnamefont {D.}~\bibnamefont {Hoyland}}, \bibinfo {author} {\bibfnamefont {A.}~\bibnamefont {Mitchell}}, \bibinfo {author} {\bibfnamefont {J.}~\bibnamefont {Van~Dongen}}, \bibinfo {author} {\bibfnamefont {C.}~\bibnamefont {Mow-Lowry}}, \bibinfo {author} {\bibfnamefont {A.}~\bibnamefont {Cumming}}, \emph {et~al.},\ }\bibfield  {title} {\bibinfo {title} {Active platform stabilization with a 6d seismometer},\ }\href@noop {} {\bibfield  {journal} {\bibinfo  {journal} {Applied Physics Letters}\ }\textbf {\bibinfo {volume} {121}},\ \bibinfo {pages} {174101} (\bibinfo {year} {2022})}\BibitemShut {NoStop}%
\bibitem [{\citenamefont {Prokhorov}\ \emph {et~al.}(2023)\citenamefont {Prokhorov}, \citenamefont {Cooper}, \citenamefont {Ubhi}, \citenamefont {Mow-Lowry}, \citenamefont {Bryant}, \citenamefont {Dmitriev}, \citenamefont {Di~Fronzo}, \citenamefont {Collins}, \citenamefont {Gill}, \citenamefont {Mitchell} \emph {et~al.}}]{prokhorov2023design}%
  \BibitemOpen
  \bibfield  {author} {\bibinfo {author} {\bibfnamefont {L.}~\bibnamefont {Prokhorov}}, \bibinfo {author} {\bibfnamefont {S.}~\bibnamefont {Cooper}}, \bibinfo {author} {\bibfnamefont {A.~S.}\ \bibnamefont {Ubhi}}, \bibinfo {author} {\bibfnamefont {C.}~\bibnamefont {Mow-Lowry}}, \bibinfo {author} {\bibfnamefont {J.}~\bibnamefont {Bryant}}, \bibinfo {author} {\bibfnamefont {A.}~\bibnamefont {Dmitriev}}, \bibinfo {author} {\bibfnamefont {C.}~\bibnamefont {Di~Fronzo}}, \bibinfo {author} {\bibfnamefont {C.~J.}\ \bibnamefont {Collins}}, \bibinfo {author} {\bibfnamefont {A.}~\bibnamefont {Gill}}, \bibinfo {author} {\bibfnamefont {A.}~\bibnamefont {Mitchell}}, \emph {et~al.},\ }\bibfield  {title} {\bibinfo {title} {Design and sensitivity of a 6-axis seismometer for gravitational wave observatories},\ }\href@noop {} {\bibfield  {journal} {\bibinfo  {journal} {arXiv preprint arXiv:2307.12891}\ } (\bibinfo {year} {2023})}\BibitemShut {NoStop}%
\bibitem [{\citenamefont {Komori}\ \emph {et~al.}(2020)\citenamefont {Komori}, \citenamefont {Enomoto}, \citenamefont {Ooi}, \citenamefont {Miyazaki}, \citenamefont {Matsumoto}, \citenamefont {Sudhir}, \citenamefont {Michimura},\ and\ \citenamefont {Ando}}]{komori2020attonewton}%
  \BibitemOpen
  \bibfield  {author} {\bibinfo {author} {\bibfnamefont {K.}~\bibnamefont {Komori}}, \bibinfo {author} {\bibfnamefont {Y.}~\bibnamefont {Enomoto}}, \bibinfo {author} {\bibfnamefont {C.~P.}\ \bibnamefont {Ooi}}, \bibinfo {author} {\bibfnamefont {Y.}~\bibnamefont {Miyazaki}}, \bibinfo {author} {\bibfnamefont {N.}~\bibnamefont {Matsumoto}}, \bibinfo {author} {\bibfnamefont {V.}~\bibnamefont {Sudhir}}, \bibinfo {author} {\bibfnamefont {Y.}~\bibnamefont {Michimura}},\ and\ \bibinfo {author} {\bibfnamefont {M.}~\bibnamefont {Ando}},\ }\bibfield  {title} {\bibinfo {title} {Attonewton-meter torque sensing with a macroscopic optomechanical torsion pendulum},\ }\href@noop {} {\bibfield  {journal} {\bibinfo  {journal} {Physical Review A}\ }\textbf {\bibinfo {volume} {101}},\ \bibinfo {pages} {011802} (\bibinfo {year} {2020})}\BibitemShut {NoStop}%
\bibitem [{\citenamefont {Chua}\ \emph {et~al.}(2023)\citenamefont {Chua}, \citenamefont {Holland}, \citenamefont {Forsyth}, \citenamefont {Kulur~Ramamohan}, \citenamefont {Zhang}, \citenamefont {Wright}, \citenamefont {Shaddock}, \citenamefont {McClelland},\ and\ \citenamefont {Slagmolen}}]{chua2023torsion}%
  \BibitemOpen
  \bibfield  {author} {\bibinfo {author} {\bibfnamefont {S.}~\bibnamefont {Chua}}, \bibinfo {author} {\bibfnamefont {N.}~\bibnamefont {Holland}}, \bibinfo {author} {\bibfnamefont {P.}~\bibnamefont {Forsyth}}, \bibinfo {author} {\bibfnamefont {A.}~\bibnamefont {Kulur~Ramamohan}}, \bibinfo {author} {\bibfnamefont {Y.}~\bibnamefont {Zhang}}, \bibinfo {author} {\bibfnamefont {J.}~\bibnamefont {Wright}}, \bibinfo {author} {\bibfnamefont {D.}~\bibnamefont {Shaddock}}, \bibinfo {author} {\bibfnamefont {D.}~\bibnamefont {McClelland}},\ and\ \bibinfo {author} {\bibfnamefont {B.}~\bibnamefont {Slagmolen}},\ }\bibfield  {title} {\bibinfo {title} {The torsion pendulum dual oscillator for low-frequency newtonian noise detection},\ }\href@noop {} {\bibfield  {journal} {\bibinfo  {journal} {Applied Physics Letters}\ }\textbf {\bibinfo {volume} {122}} (\bibinfo {year} {2023})}\BibitemShut {NoStop}%
\bibitem [{\citenamefont {Miao}\ \emph {et~al.}(2020)\citenamefont {Miao}, \citenamefont {Martynov}, \citenamefont {Yang},\ and\ \citenamefont {Datta}}]{miao2020quantum}%
  \BibitemOpen
  \bibfield  {author} {\bibinfo {author} {\bibfnamefont {H.}~\bibnamefont {Miao}}, \bibinfo {author} {\bibfnamefont {D.}~\bibnamefont {Martynov}}, \bibinfo {author} {\bibfnamefont {H.}~\bibnamefont {Yang}},\ and\ \bibinfo {author} {\bibfnamefont {A.}~\bibnamefont {Datta}},\ }\bibfield  {title} {\bibinfo {title} {Quantum correlations of light mediated by gravity},\ }\href@noop {} {\bibfield  {journal} {\bibinfo  {journal} {Physical Review A}\ }\textbf {\bibinfo {volume} {101}},\ \bibinfo {pages} {063804} (\bibinfo {year} {2020})}\BibitemShut {NoStop}%
\bibitem [{\citenamefont {Plissi}\ \emph {et~al.}(2004)\citenamefont {Plissi}, \citenamefont {Torrie}, \citenamefont {Barton}, \citenamefont {Robertson}, \citenamefont {Grant}, \citenamefont {Cantley}, \citenamefont {Strain}, \citenamefont {Willems}, \citenamefont {Romie}, \citenamefont {Skeldon} \emph {et~al.}}]{plissi2004investigation}%
  \BibitemOpen
  \bibfield  {author} {\bibinfo {author} {\bibfnamefont {M.}~\bibnamefont {Plissi}}, \bibinfo {author} {\bibfnamefont {C.}~\bibnamefont {Torrie}}, \bibinfo {author} {\bibfnamefont {M.}~\bibnamefont {Barton}}, \bibinfo {author} {\bibfnamefont {N.}~\bibnamefont {Robertson}}, \bibinfo {author} {\bibfnamefont {A.}~\bibnamefont {Grant}}, \bibinfo {author} {\bibfnamefont {C.}~\bibnamefont {Cantley}}, \bibinfo {author} {\bibfnamefont {K.}~\bibnamefont {Strain}}, \bibinfo {author} {\bibfnamefont {P.}~\bibnamefont {Willems}}, \bibinfo {author} {\bibfnamefont {J.}~\bibnamefont {Romie}}, \bibinfo {author} {\bibfnamefont {K.}~\bibnamefont {Skeldon}}, \emph {et~al.},\ }\bibfield  {title} {\bibinfo {title} {An investigation of eddy-current damping of multi-stage pendulum suspensions for use in interferometric gravitational wave detectors},\ }\href@noop {} {\bibfield  {journal} {\bibinfo  {journal} {Review of scientific instruments}\ }\textbf {\bibinfo {volume} {75}},\ \bibinfo {pages} {4516} (\bibinfo {year}
  {2004})}\BibitemShut {NoStop}%
\bibitem [{\citenamefont {Black}(2001)}]{Black_PoundDreverHall_2001}%
  \BibitemOpen
  \bibfield  {author} {\bibinfo {author} {\bibfnamefont {E.~D.}\ \bibnamefont {Black}},\ }\bibfield  {title} {\bibinfo {title} {{An introduction to Pound-Drever-Hall laser frequency stabilization}},\ }\bibfield  {journal} {\bibinfo  {journal} {Am. J. Phys.}\ }\href {https://doi.org/10.1119/1.1286663} {10.1119/1.1286663} (\bibinfo {year} {2001})\BibitemShut {NoStop}%
\bibitem [{\citenamefont {Smetana}\ \emph {et~al.}(2023)\citenamefont {Smetana}, \citenamefont {Yan}, \citenamefont {Boyer},\ and\ \citenamefont {Martynov}}]{smetana2023reaching}%
  \BibitemOpen
  \bibfield  {author} {\bibinfo {author} {\bibfnamefont {J.}~\bibnamefont {Smetana}}, \bibinfo {author} {\bibfnamefont {T.}~\bibnamefont {Yan}}, \bibinfo {author} {\bibfnamefont {V.}~\bibnamefont {Boyer}},\ and\ \bibinfo {author} {\bibfnamefont {D.}~\bibnamefont {Martynov}},\ }\bibfield  {title} {\bibinfo {title} {Reaching a macroscopic quantum state in a suspended interferometer},\ }in\ \href@noop {} {\emph {\bibinfo {booktitle} {Quantum Technology: Driving Commercialisation of an Enabling Science III}}},\ Vol.\ \bibinfo {volume} {12335}\ (\bibinfo {organization} {SPIE},\ \bibinfo {year} {2023})\ pp.\ \bibinfo {pages} {21--30}\BibitemShut {NoStop}%
\bibitem [{\citenamefont {Smetana}\ \emph {et~al.}(2024)\citenamefont {Smetana}, \citenamefont {Yan}, \citenamefont {Boyer},\ and\ \citenamefont {Martynov}}]{smetana2024high}%
  \BibitemOpen
  \bibfield  {author} {\bibinfo {author} {\bibfnamefont {J.}~\bibnamefont {Smetana}}, \bibinfo {author} {\bibfnamefont {T.}~\bibnamefont {Yan}}, \bibinfo {author} {\bibfnamefont {V.}~\bibnamefont {Boyer}},\ and\ \bibinfo {author} {\bibfnamefont {D.}~\bibnamefont {Martynov}},\ }\bibfield  {title} {\bibinfo {title} {A high-finesse suspended interferometric sensor for macroscopic quantum mechanics with femtometre sensitivity},\ }\href@noop {} {\bibfield  {journal} {\bibinfo  {journal} {Sensors}\ }\textbf {\bibinfo {volume} {24}},\ \bibinfo {pages} {2375} (\bibinfo {year} {2024})}\BibitemShut {NoStop}%
\bibitem [{\citenamefont {Carbone}\ \emph {et~al.}(2012)\citenamefont {Carbone}, \citenamefont {Aston}, \citenamefont {Cutler}, \citenamefont {Freise}, \citenamefont {Greenhalgh}, \citenamefont {Heefner}, \citenamefont {Hoyland}, \citenamefont {Lockerbie}, \citenamefont {Lodhia}, \citenamefont {Robertson} \emph {et~al.}}]{carbone2012sensors}%
  \BibitemOpen
  \bibfield  {author} {\bibinfo {author} {\bibfnamefont {L.}~\bibnamefont {Carbone}}, \bibinfo {author} {\bibfnamefont {S.}~\bibnamefont {Aston}}, \bibinfo {author} {\bibfnamefont {R.}~\bibnamefont {Cutler}}, \bibinfo {author} {\bibfnamefont {A.}~\bibnamefont {Freise}}, \bibinfo {author} {\bibfnamefont {J.}~\bibnamefont {Greenhalgh}}, \bibinfo {author} {\bibfnamefont {J.}~\bibnamefont {Heefner}}, \bibinfo {author} {\bibfnamefont {D.}~\bibnamefont {Hoyland}}, \bibinfo {author} {\bibfnamefont {N.}~\bibnamefont {Lockerbie}}, \bibinfo {author} {\bibfnamefont {D.}~\bibnamefont {Lodhia}}, \bibinfo {author} {\bibfnamefont {N.}~\bibnamefont {Robertson}}, \emph {et~al.},\ }\bibfield  {title} {\bibinfo {title} {Sensors and actuators for the advanced ligo mirror suspensions},\ }\href@noop {} {\bibfield  {journal} {\bibinfo  {journal} {Classical and Quantum Gravity}\ }\textbf {\bibinfo {volume} {29}},\ \bibinfo {pages} {115005} (\bibinfo {year} {2012})}\BibitemShut {NoStop}%
\bibitem [{\citenamefont {Smetana}\ \emph {et~al.}(2022)\citenamefont {Smetana}, \citenamefont {Walters}, \citenamefont {Bauchinger}, \citenamefont {Ubhi}, \citenamefont {Cooper}, \citenamefont {Hoyland}, \citenamefont {Abbott}, \citenamefont {Baune}, \citenamefont {Fritchel}, \citenamefont {Gerberding} \emph {et~al.}}]{smetana2022compact}%
  \BibitemOpen
  \bibfield  {author} {\bibinfo {author} {\bibfnamefont {J.}~\bibnamefont {Smetana}}, \bibinfo {author} {\bibfnamefont {R.}~\bibnamefont {Walters}}, \bibinfo {author} {\bibfnamefont {S.}~\bibnamefont {Bauchinger}}, \bibinfo {author} {\bibfnamefont {A.~S.}\ \bibnamefont {Ubhi}}, \bibinfo {author} {\bibfnamefont {S.}~\bibnamefont {Cooper}}, \bibinfo {author} {\bibfnamefont {D.}~\bibnamefont {Hoyland}}, \bibinfo {author} {\bibfnamefont {R.}~\bibnamefont {Abbott}}, \bibinfo {author} {\bibfnamefont {C.}~\bibnamefont {Baune}}, \bibinfo {author} {\bibfnamefont {P.}~\bibnamefont {Fritchel}}, \bibinfo {author} {\bibfnamefont {O.}~\bibnamefont {Gerberding}}, \emph {et~al.},\ }\bibfield  {title} {\bibinfo {title} {Compact michelson interferometers with subpicometer sensitivity},\ }\href@noop {} {\bibfield  {journal} {\bibinfo  {journal} {Physical Review Applied}\ }\textbf {\bibinfo {volume} {18}},\ \bibinfo {pages} {034040} (\bibinfo {year} {2022})}\BibitemShut {NoStop}%
\bibitem [{\citenamefont {Robertson}\ \emph {et~al.}(2009)\citenamefont {Robertson}, \citenamefont {Hough},\ and\ \citenamefont {Collaboration}}]{robertson2009gas}%
  \BibitemOpen
  \bibfield  {author} {\bibinfo {author} {\bibfnamefont {N.~A.}\ \bibnamefont {Robertson}}, \bibinfo {author} {\bibfnamefont {J.}~\bibnamefont {Hough}},\ and\ \bibinfo {author} {\bibfnamefont {L.~S.}\ \bibnamefont {Collaboration}},\ }\bibfield  {title} {\bibinfo {title} {Gas damping in advanced ligo suspensions},\ }\href@noop {} {\bibfield  {journal} {\bibinfo  {journal} {LIGO Document: LIGO-T0900416-v2}\ } (\bibinfo {year} {2009})}\BibitemShut {NoStop}%
\bibitem [{\citenamefont {Cumming}\ \emph {et~al.}(2009)\citenamefont {Cumming}, \citenamefont {Heptonstall}, \citenamefont {Kumar}, \citenamefont {Cunningham}, \citenamefont {Torrie}, \citenamefont {Barton}, \citenamefont {Strain}, \citenamefont {Hough},\ and\ \citenamefont {Rowan}}]{cumming2009finite}%
  \BibitemOpen
  \bibfield  {author} {\bibinfo {author} {\bibfnamefont {A.}~\bibnamefont {Cumming}}, \bibinfo {author} {\bibfnamefont {A.}~\bibnamefont {Heptonstall}}, \bibinfo {author} {\bibfnamefont {R.}~\bibnamefont {Kumar}}, \bibinfo {author} {\bibfnamefont {W.}~\bibnamefont {Cunningham}}, \bibinfo {author} {\bibfnamefont {C.}~\bibnamefont {Torrie}}, \bibinfo {author} {\bibfnamefont {M.}~\bibnamefont {Barton}}, \bibinfo {author} {\bibfnamefont {K.}~\bibnamefont {Strain}}, \bibinfo {author} {\bibfnamefont {J.}~\bibnamefont {Hough}},\ and\ \bibinfo {author} {\bibfnamefont {S.}~\bibnamefont {Rowan}},\ }\bibfield  {title} {\bibinfo {title} {Finite element modelling of the mechanical loss of silica suspension fibres for advanced gravitational wave detectors},\ }\href@noop {} {\bibfield  {journal} {\bibinfo  {journal} {Classical and Quantum Gravity}\ }\textbf {\bibinfo {volume} {26}},\ \bibinfo {pages} {215012} (\bibinfo {year} {2009})}\BibitemShut {NoStop}%
\bibitem [{\citenamefont {Cumming}\ \emph {et~al.}(2012)\citenamefont {Cumming}, \citenamefont {Bell}, \citenamefont {Barsotti}, \citenamefont {Barton}, \citenamefont {Cagnoli}, \citenamefont {Cook}, \citenamefont {Cunningham}, \citenamefont {Evans}, \citenamefont {Hammond}, \citenamefont {Harry} \emph {et~al.}}]{cumming2012design}%
  \BibitemOpen
  \bibfield  {author} {\bibinfo {author} {\bibfnamefont {A.}~\bibnamefont {Cumming}}, \bibinfo {author} {\bibfnamefont {A.}~\bibnamefont {Bell}}, \bibinfo {author} {\bibfnamefont {L.}~\bibnamefont {Barsotti}}, \bibinfo {author} {\bibfnamefont {M.}~\bibnamefont {Barton}}, \bibinfo {author} {\bibfnamefont {G.}~\bibnamefont {Cagnoli}}, \bibinfo {author} {\bibfnamefont {D.}~\bibnamefont {Cook}}, \bibinfo {author} {\bibfnamefont {L.}~\bibnamefont {Cunningham}}, \bibinfo {author} {\bibfnamefont {M.}~\bibnamefont {Evans}}, \bibinfo {author} {\bibfnamefont {G.}~\bibnamefont {Hammond}}, \bibinfo {author} {\bibfnamefont {G.}~\bibnamefont {Harry}}, \emph {et~al.},\ }\bibfield  {title} {\bibinfo {title} {Design and development of the advanced ligo monolithic fused silica suspension},\ }\href@noop {} {\bibfield  {journal} {\bibinfo  {journal} {Classical and Quantum Gravity}\ }\textbf {\bibinfo {volume} {29}},\ \bibinfo {pages} {035003} (\bibinfo {year} {2012})}\BibitemShut {NoStop}%
\bibitem [{\citenamefont {Willems}\ \emph {et~al.}(2002)\citenamefont {Willems}, \citenamefont {Sannibale}, \citenamefont {Weel},\ and\ \citenamefont {Mitrofanov}}]{willems2002investigations}%
  \BibitemOpen
  \bibfield  {author} {\bibinfo {author} {\bibfnamefont {P.}~\bibnamefont {Willems}}, \bibinfo {author} {\bibfnamefont {V.}~\bibnamefont {Sannibale}}, \bibinfo {author} {\bibfnamefont {J.}~\bibnamefont {Weel}},\ and\ \bibinfo {author} {\bibfnamefont {V.}~\bibnamefont {Mitrofanov}},\ }\bibfield  {title} {\bibinfo {title} {Investigations of the dynamics and mechanical dissipation of a fused silica suspension},\ }\href@noop {} {\bibfield  {journal} {\bibinfo  {journal} {Physics Letters A}\ }\textbf {\bibinfo {volume} {297}},\ \bibinfo {pages} {37} (\bibinfo {year} {2002})}\BibitemShut {NoStop}%
\bibitem [{\citenamefont {Gretarsson}\ and\ \citenamefont {Harry}(1999)}]{gretarsson1999dissipation}%
  \BibitemOpen
  \bibfield  {author} {\bibinfo {author} {\bibfnamefont {A.~M.}\ \bibnamefont {Gretarsson}}\ and\ \bibinfo {author} {\bibfnamefont {G.~M.}\ \bibnamefont {Harry}},\ }\bibfield  {title} {\bibinfo {title} {Dissipation of mechanical energy in fused silica fibers},\ }\href@noop {} {\bibfield  {journal} {\bibinfo  {journal} {Review of scientific instruments}\ }\textbf {\bibinfo {volume} {70}},\ \bibinfo {pages} {4081} (\bibinfo {year} {1999})}\BibitemShut {NoStop}%
\end{thebibliography}%
\appendix

\section{Estimation of the mechanical quality factor}
\label{appendix:Q-factor}
We estimate the mechanical quality factor $Q$ from different sources and conclude that the $Q$ factor of our torsion mode is at least \(50000\). They are detailed as follows.

\subsection{Quality factor estimated from DAC noise}
We measured the torsion mode's ringdown over 90 hours. No visible ringdown was observed due to the high mechanical quality factor. However, we detected RMS rised from \SI{4}{\um} to \SI{7}{\um} in the envelope of the ringdown caused by the actuating noise of the coil originating from the DAC (Digital to analog convert) noise. We estimate \SI{5e4}{} from RMS $= \SI{4}{\um} \times \sqrt{ Q f} =$ \SI{7}{\um}.

\subsection{Gas damping}
% \section{Introduction}
From \cite{robertson2009gas}, we have the mechanical quality factor contributed from the gas damping as follows,
\begin{equation}
\begin{aligned}
Q = \left(\frac{\pi}{2} \right)^{\frac{3}{2}} \rho H f \sqrt{\frac{RT}{M_{\rm m}}} \frac{1}{P},
\end{aligned}
\end{equation}
where \(\rho\) is the specific mass, $H$ is the thickness, $R$ is the ideal gas constant, $T$ is the temperature, $f$ is the oscillating frequency, \(M_{\rm m}\) is the molar weight of the gas and $P$ is the pressure. 
In our case, we turn \(\rho H\) into \(M/A\), where $M$ is the mass, $A$ is the surface area. In our case, \(M = \SI{1}{\kg}\), \(A = \SI{0.05}{\m} \times \SI{0.5}{\m} \times 3\) for the mass, \(M_{\rm m} = 2\times 10^{-3}\) for H\textsubscript{2}, and \(P = \SI{2e-6}{Torr}\). So we have the $Q$-factor contributed from gas damping is \SI{6.6e4}{}.

\subsection{Fibre thermoelastic}
Thermoelastic loss is \cite{cumming2009finite, cumming2012design},
\begin{equation}
\begin{aligned}
\Phi_{\rm thermoelastic}(\omega) = \frac{YT}{\rho C} \left( \alpha - \sigma_0 \frac{\beta}{Y} \right) ^2 \left( \frac{\omega \tau}{1+(\omega \tau)^2}\right),
\end{aligned}
\end{equation}
with
\begin{equation}
\begin{aligned}
\tau = \frac{1}{4.32 \pi} \frac{\rho C d^2}{\kappa},
\end{aligned}
\end{equation}
where $Y$ is Young’s modulus of the fiber, $C$ is the specific heat capacity of the material per unit mass, \(\kappa\) is its thermal conductivity, \(\rho\) is the material density, \(\alpha\) is the coefficient of linear thermal expansion, \(\sigma_0\) is the static stress in the fiber due to the suspended load, \(\beta = (1/Y) (dY/dT)\) is the thermal elastic coefficient and d is the diameter of the fiber.

As the material of the fiber is fused silica, and the diameter of the fiber is \SI{100}{\um}, with \(\beta = \SI{1.52e-4}{\per\K}\) taken from Willems \cite{willems2002investigations}, we calculated the \(\Phi_{\rm thermoelastic} \sim 10^{-12}\) at the resonance frequency. The loss mechanism is negligible in our mechanical system.

\subsection{Fibre surface}
The fibre surface loss is given by \cite{cumming2012design},
\begin{equation}
\begin{aligned}
\Phi_{\rm surface} \approx \frac{8 h \Phi_s}{d},
\end{aligned}
\end{equation}
where \( h \Phi_s \) is the product of the mechanical loss of the material surface, \( \Phi_s \), and the depth, $h$, over which surface loss mechanisms are believed to occur. Here, \( h \Phi_{s} = \SI{6.15e-12}{\m}\) was taken from Gretarsson \cite{gretarsson1999dissipation}. So we calculated \(Q_{\rm surface} = 1/\Phi_{\rm surface} = \SI{2.0e6}{}\).

\subsection{Eddy current damping}
We have a horizontal coil-magnet actuator for the torsion mode control. Two \SI{1}{\mm^3} SmCo magnets are mounted with a separation of \SI{2}{mm}, with their poles in opposite directions. The coil is located \SI{10}{mm} from the closest magnet, and the central axes of the coil and the magnets coincide. Simulating with Comsol, we got the damping factor \( \gamma = \SI{5.3e-9}{\N \cdot \s \per \m}\). So the $Q$ factor contributed from eddy current damping is \(Q_{\rm eddy current} = (\omega_{\rm m} I)/(\gamma a^2) = 4.0 \times 10^5\), where $I$ is the moment of inertia, a is the distance from the magnet to the suspension point.

\begin{figure*}
\includegraphics[width=0.8\textwidth]{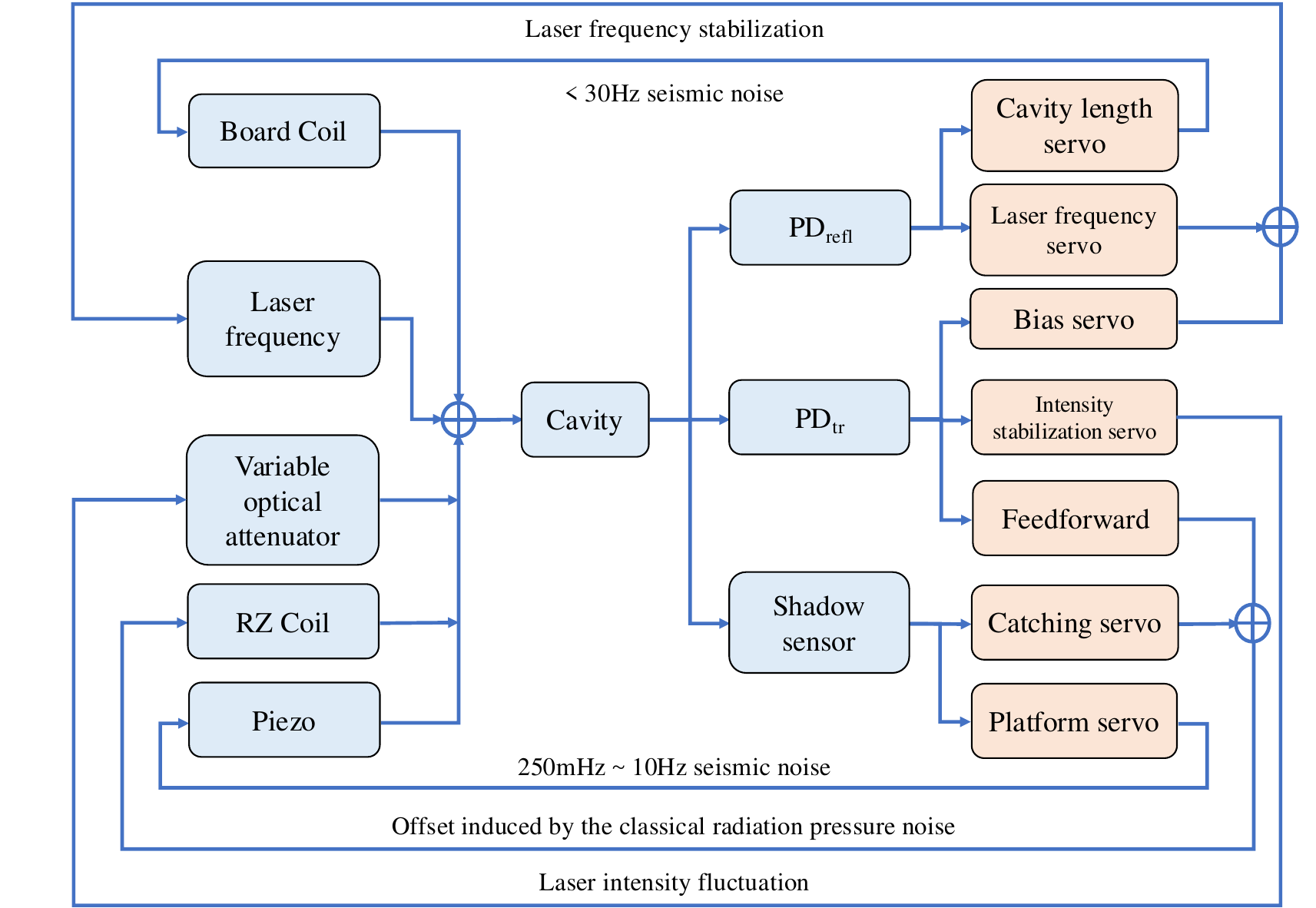}% Here is how to import EPS art
\caption{\label{fig:epsart2} Control architecture for cavity stabilization. Seven servos were utilized for locking, with laser frequency stabilized using the Pound-Drever-Hall method. The bias servo is engaged for removing the DC offset in the error signal. Seismic noise below \SI{30}{\Hz} was mitigated by platform stabilization from \SI{250}{\mHz} to \SI{10}{\Hz}, and by modulating cavity length with a bandwidth of \SI{30}{\Hz}. The classical radiation pressure force was compensated by a feed-forward and a catching servo, while the intensity noise was controlled via an intensity stabilization servo. %TY: sorted \textcolor{red}{DM: I think this figure should move to the appendix.}
}
\end{figure*}

\section{The control architecture}
\label{appendix:ctrl}
The control architecture illustrated in Figure~\ref{fig:epsart2} was designed to address the unique challenges of a high-finesse cavity suspended on a torsion pendulum, with a torsional resonance frequency of 0.6mHz. This system is highly sensitive to seismic noise, radiation pressure, and laser intensity fluctuations. To ensure robust stabilization over a broad frequency range, the architecture integrates multiple feedback and feed-forward strategies to maintain cavity resonance effectively.

The architecture comprises three main components: sensing, feedback and feedforward control, and actuation. The sensing layer collects real-time data from shadow sensors and photodiodes, which is processed by the feedback and feedforward control layer to generate precise control signals. These signals are then transmitted to the actuators, ensuring the cavity remains stable and aligned.

\section{The optical spring effect}
\label{appendix:opticalspring}
Imagine a cavity with one mirror fixed and one mirror movable. The amplitude fluctuation of the input laser is \(a_{\rm in}\). \(\mathcal{T}\) is the Transmittance, \(r\) is the reflection coefficient of the cavity mirror. \(\omega_{\rm m}\) is the mechanical frequency of the movable mirror. \(x\) is the displacement of the movable mirror, \(t\) is the time. \(A_c\) is the circulating power of $A$ in the cavity, while \(A_{\rm in}(t)\) is the input power, \(t_{\rm rt}\) is the round-trip time delay, \(\Delta\) is the round trip phase of detuning. In our case, \(\Delta=\SI{8e-9}{}\), \(t_{\rm rt} = \SI{6.4e-10}{\s}\).

For the fluctuation, we have
\begin{equation}
\begin{aligned}
a_c(t) &= \sqrt{\mathcal{T}} a_{\rm in}(t) + r^2 a_c (t-t_{\rm rt}) e^{i \Delta} + \\
&+A_c r^2 e^{i \frac{2 \pi}{\lambda} 2 x (t-t_{\rm rt}/2)} e^{i \Delta},
\end{aligned}
\end{equation}
where \(A_c r e^{i \frac{2 \pi}{\lambda} 2 x (t-t_{\rm rt}/2)} e^{i \Delta /2}\) is due to the modulation from the movable mirror, which introduces fluctuations to the amplitude. \(2 x (t-t_{\rm rt}/2)\) is the round trip optical path difference, \((2 \pi/\lambda) 2 x (t-t_{\rm rt}/2)\) is the corresponding phase. 

In this experiment, \(\mathcal{T} = 8 \times 10^{-6}\), we can ignore the term \(\sqrt{\mathcal{T}} a_{\rm in}(t)\). Also, as we are interested in the mHz level, \(t_{\rm rt}\) can be ignored. After some simplification, we have
\begin{equation}
\begin{aligned}
a_c(t) = r^2 a_c(t) e^{i \Delta}+  \frac{4\pi ix(t)}{\lambda G_{\rm FSS}} A_c e^{i \Delta},
\end{aligned}
\end{equation}
and then
\begin{equation} \label{eq:a}
\begin{aligned}
a_c(t) \approx \frac{i 4 \pi A_c x(t)}{\lambda \mathcal{T} G_{\rm FSS}} (1+i \frac{\Delta}{\mathcal{T}}).
\end{aligned}
\end{equation}

For a torsion pendulum, the RZ dynamics is: \(x(t)=\alpha(t)L\), where \(L=\SI{0.6}{\m}\) is the length of the mass. In that case, considering the force applied to this harmonic torsion pendulum, we have:

\begin{equation} 
\begin{aligned}
I_{\rm RZ} (\ddot{\alpha} + \gamma \dot{\alpha} + \omega_m^2 \alpha) = \left(\frac{2}{cG_{\rm ISS}} A(a +a^*)+F_{\rm ext}\right) L,
\end{aligned}
\end{equation}
where \(G_{\rm ISS}\) is the gain of the intensity stabilization servo. Applying Eq.\ref{eq:a}, we have

\begin{equation}
\begin{aligned}
I_{\rm RZ} (\ddot{\alpha} + \gamma \dot{\alpha} + \omega_m^2 \alpha) = -\frac{8\pi P_0 L^2 }{\lambda \mathcal{T} G_{\rm FSS}}\frac{\Delta}{\mathcal{T}} \frac{2}{cG_{\rm ISS}} \alpha +L F_{\rm ext},
\end{aligned}
\end{equation}
where \(P_0 = \SI{80}{\W}\) is the power circulating in the cavity, rearranging the equation leads to:

\begin{equation}
\begin{aligned}
&I_{\rm RZ} \left(\ddot{\alpha} + \gamma \dot{\alpha} + \omega_m^2 \left( 1+\frac{16\pi P_0 L^2}{I_{\rm RZ}\lambda \mathcal{T} c G_{\rm FSS}G_{\rm ISS} \omega_m^2} \frac{\Delta}{\mathcal{T}}\right) \alpha \right) \\
 & = F_{\rm ext} L.
\end{aligned}
\end{equation}

The shifted mechanical frequency can be obtained from the above equation: 
\begin{equation}
\omega^2_{\rm ms}=\omega_m^2 \left[1+\frac{16\pi P_0 L^2}{I_{\rm RZ}\lambda \mathcal{T} c G_{\rm FSS}G_{\rm ISS} \omega_m^2}\frac{\Delta}{\mathcal{T}}\right].
\end{equation}
In our case, \(G_{\rm ISS} = 100\) when \(f< \SI{0.1}{Hz}\), \(G_{\rm FSS} \approx 10^{11}\) when \(f\leq \SI{0.1}{\mHz}\), hence the relative modification of the mechanical resonance frequency is 
\begin{equation}
\frac{\omega_{\rm ms}^2-\omega_m^2}{\omega_m^2}\approx 2.0\times 10 ^{-5},
\end{equation}
which corresponds to  \SI{5.9}{nHz} frequency shift and is too small to be resolved during our observing run.

\section{Comparison of the shadow sensor signal and cavity control signal}
\label{appendix:SN_RZ}
\begin{figure*}
\includegraphics[width=0.7\textwidth]{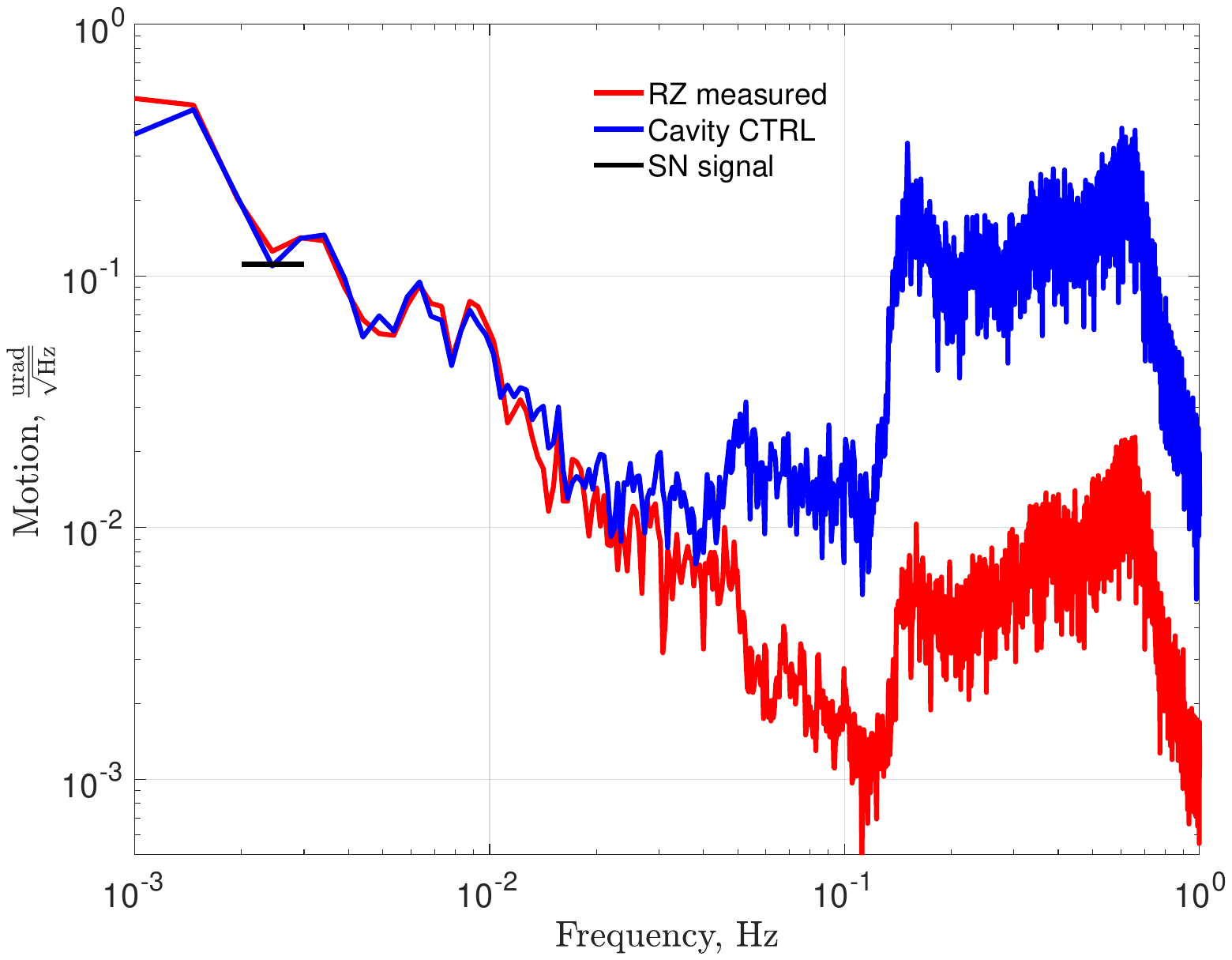}% Here is how to import EPS art
\caption{\label{fig:SN_RZ} Comparison of the RZ measured signal (shadow sensor signal) and the cavity control signal during one of the engineering runs. Below 20 mHz, the cavity control signal remains coherent with the shadow sensor signal, indicating that both capture the torsional motion of the system. At higher frequencies, the cavity control signal exceeds the shadow sensor signal due to the coupling of longitudinal seismic noise into the cavity response. The black line corresponds to the expected SN signal.
}
\end{figure*}

As discussed in the main text, the cavity readout noise is significantly lower than the shadow sensor noise. However, at the frequency of interest, the system remains limited by actuator force noise around \SI{1}{mHz}. To illustrate the relative behavior of the two readout signals, we present a comparison between the RZ measured signal (shadow sensor signal) and the cavity control signal in Figure~\ref{fig:SN_RZ}.

Below 20 mHz, the cavity control signal remains coherent with the shadow sensor signal, confirming that both capture the torsional motion of the system. At higher frequencies, the cavity control signal exceeds the shadow sensor signal due to the coupling of longitudinal seismic noise into the cavity response. This comparison highlights the practical limitations of the cavity readout in this setup and supports the choice of using the shadow sensor for signal extraction in this frequency range.

\section{The uncertainty of the macroscopic motion}
\label{appendix:macroscopic motion}
In the quadratic SN potential region, under the preselection quantum measurement prescription, thermal noise is treated as classical. All other noises, except for quantum radiation pressure noise, are also assumed to be classical. We now proceed to estimate the quantum uncertainty of the torsion pendulum specifically due to quantum radiation pressure noise.
The equations of motion of the torsion pendulum can be written as:
\begin{equation}
I_{\rm RZ}\ddot{\Theta}=-I_{\rm RZ}\omega_q^2\Theta-\gamma_mI_{\rm RZ}\dot{\Theta}+\hbar\alpha L\hat{a}_1+I_{\rm RZ}\omega_{\rm SN}\langle\Theta\rangle,
\end{equation}
thereby
\begin{equation}
\Theta(\Omega)=\frac{\hbar\alpha L\hat{a}_1+I_{\rm RZ}\omega_{\rm SN}\langle\Theta\rangle}{I_{\rm RZ}(\omega_q^2-\Omega^2-i\gamma_m\Omega)},
\end{equation}
where \(\omega_q = \sqrt{\omega_m^2 + \omega_{\text{SN}}^2}\), \(\omega_{\text{SN}} =2\pi\times \SI{2.93}{\rm rad/s}\). Additionally, \(\alpha = \sqrt{8\omega_m P_{\rm cav}/\hbar T c^2}\), we choose optical wavelength \(\lambda = \SI{1550}{
nm}\), \(T = \SI{8}{\ppm}\), and the cavity light power is \(P_{\text{cav}} = \SI{80}{\W}\). The moment of inertia of the RZ is \(I_{\text{RZ}} = \SI{0.14}{\kg\cdot \m^2}\), and the length of the torsion pendulum is \(L = \SI{0.6}{\m}\). Then the noise spectrum of rotation due to the radiation pressure is:
\begin{equation}
S^{\rm rad}_{\Theta\Theta}(\Omega)=\frac{\hbar^2\alpha^2L^2}{I_{\rm RZ}^2|\omega_q^2-\Omega^2-i\Omega\gamma_m|^2}.
\end{equation}
The quantum uncertainty of the torsion pendulum can be estimated as,
\begin{equation}
\Delta\Theta\approx\sqrt{S^{\rm rad}_{\Theta\Theta}(\omega_q)\gamma_m}=\sqrt{\frac{\hbar^2\alpha^2L^2}{I_{\rm RZ}^2\omega_q^2\gamma_m}}.
\end{equation}
Using the eigenfrequency $\omega_m=2\pi \times 0.6$\,mHz and $Q=\SI{5e4}{}$, the result is evaluated as,
\begin{equation}
\Delta\Theta=1.37\times10^{-8}\,{\rm rad.}
\end{equation}
Suppose the atom is situated at the distance $d=0.6$\,m from the rotation center, the uncertainty becomes $\Delta X=L\Delta\Theta=\SI{8.24e-9}{}$\,m, which is larger than $\Delta x_{\rm int}$.

In conducting the experiment, a feedback loop is employed to mitigate the uncertainty in the motion of the torsion pendulum. The uncertainty of the torsion pendulum with the feedback loop is calculated as follows. The feedback servo function reads
\begin{equation}
C(s)=\frac{A_0(s+2\pi\times 10^{-3})^2}{s(s+2\pi)},
\end{equation}
and the response function of the torsion pendulum is,
\begin{equation}
\chi_{m/q}(\Omega)=\frac{1}{I_z(\omega_{m/q}^2-\Omega^2-i\gamma_m\Omega)}
\end{equation}
where $\chi_{m/q}(\Omega)$ is the response function to classical/quantum noise. The coefficient $A_0=0.038$ is obtained via  the condition $|C(\Omega)\chi_m(\Omega)|=1$ with $\Omega=2\pi\times7\times10^{-3}\,{\rm rad/s}$. In the following, we will use the kernel function $H_q(\Omega)$ defined as
\begin{equation}
H_q(\Omega)=C(\Omega)\chi_q(\Omega).
\end{equation}

In this case, the quantum motion of the torsion pendulum is presented by
\begin{equation}
\Theta(\Omega)=\frac{\chi_q(\Omega)}{1+H_q(\Omega)}\hbar\alpha L\hat{a}_1-\frac{H_q(\Omega)}{1+H_q(\Omega)}n,
\end{equation}
where $n$ is the sensor noise.  Then the spectrum of the rotational motion is:
\begin{equation}\label{eq:S_withfeedback}
S_{\Theta\Theta}(\Omega)=\frac{|\chi_q(\Omega)|^2\hbar^2\alpha^2 L^2}{|1+H_q(\Omega)|^2}+\frac{|H_q(\Omega)|^2}{|1+H_q(\Omega)|^2}S_{nn}(\Omega).
\end{equation}
Integrating the above spectrum in the frequency domain, we find that the feedback system reduces the radiation-pressure-driven uncertainty to \SI{1.21e-11}{\m}. Therefore, the uncertainty of the rotational motion is dominated by the sensor noise at the level of $\SI{3.41e-9}{\m}\gg\Delta x_{\rm int}=\SI{1.04e-11}{\m}$. This uncertainty also makes the quadratic form of SN self-gravity potential invalid.

\section{The feedback loop in the nonlinear quantum mechanics}\label{sec:feedback_role}
The influence of the feedback control process on the outgoing field spectrum is subtle in the SN gravity theory due to its nonlinearity, which can not be trivially removed.  Specifically, in the SN theory with quadratic self-gravity potential, the dynamical equation of the test mass in SN quadrature potential reads,
\begin{equation}
\begin{split}
\dot{\hat{x}}&=\frac{\hat{p}}{M},\\
\dot{\hat{p}}&=-M\omega_{\rm m}^2\hat{x}-\gamma_{\rm m}\hat{p}+\hbar\alpha\hat{a}_1+F_{\rm th}-M\omega_{\rm SN}^2(\hat{x}-\langle\hat{x}\rangle)\\&-\int_0^tC(t-t')(\hat{x}(t')+\hat{n}(t'))dt'.
\end{split}
\end{equation}
Under the pre-selection prescription\,\cite{Helou2017}, the thermal noise is separated into the quantum part and the classical part. Following a similar prescription for treating the thermal noise, we can also separate the sensor noise into the quantum part and the classical part, $\hat{n}=n_{\rm cl}+\hat{n}_{\rm zp}$, and the expectation value of the sensor noise can be written as:
\begin{equation}
\langle\hat{n}\rangle=n_{\rm cl}+\langle\hat{n}_{\rm zp}\rangle.
\end{equation}

In the frequency domain, the equations of motion are:
\begin{equation}\label{eq:eom_frequency}
\begin{split}
&\hat{x}(\Omega)=\frac{\hbar\alpha\hat{a}_1(\Omega)+F_{\rm th}(\Omega)-C(\Omega)\hat{n}(\Omega)+M\omega_{\rm SN}^2\langle\hat{x}(\Omega)\rangle}{M(\omega_{\rm q}^2-\Omega^2-i\Omega\gamma_{\rm m})+C(\Omega)},\\
&\langle\hat{x}(\Omega)\rangle=\frac{\hbar\alpha\langle\hat{a}_1(\Omega)\rangle+\langle F_{\rm th}(\Omega)\rangle-C(\Omega)\langle\hat{n}(\Omega)\rangle}{M(\omega_{\rm m}^2-\Omega^2-i\Omega\gamma_{\rm m})+C(\Omega)}.
\end{split}
\end{equation}
The solution of the above equations is:
\begin{equation}
\begin{split}
\hat{x}(\Omega)
&=\frac{\hbar\alpha\hat{a}_1(\Omega)}{M(\omega_{\rm q}^2-\Omega^2-i\Omega\gamma_{\rm m})+C(\Omega)}\\
&+\frac{F^{\rm cl}_{\rm th}(\Omega)-C(\Omega)n_{\rm cl}(\Omega)}{M(\omega_{\rm m}^2-\Omega^2-i\Omega\gamma_{\rm m})+C(\Omega)}\\&+\Delta G(\Omega)\langle\hat{B}(\Omega)\rangle,
\end{split}
\end{equation}
where 
\begin{equation}
\begin{split}
\hat{B}(\Omega)&=\hbar\alpha\hat{a}_1(\Omega)-C(\Omega)\hat{n}_{\rm zp}(\Omega)+F^{\rm zp}_{\rm th}(\Omega),\\
\Delta G(\Omega)&=\frac{1}{M(\omega_{\rm m}^2-\Omega^2-i\Omega\gamma_{\rm m})+C(\Omega)}\\&-\frac{1}{M(\omega_{\rm q}^2-\Omega^2-i\Omega\gamma_{\rm m})+C(\Omega)}.
\end{split}
\end{equation}\\

Rearranging the terms of the above equation leads to:
\begin{equation}
\begin{split}
&\hat{x}(\Omega)=\\
&\frac{\chi_q(\Omega)\hbar\alpha\hat{a}_1(\Omega)+\chi_q(\Omega)F^{\rm zp}_{\rm th}(\Omega)-C(\Omega)\chi_q(\Omega)\hat{n}_{\rm zp}(\Omega)}{1+\chi_q(\Omega)C(\Omega)}\\
&+\frac{\chi_m(\Omega)F_{\rm th}^{\rm cl}(\Omega)-C(\Omega)\chi_m(\Omega)n_{\rm cl}(\Omega)}{1+\chi_m(\Omega)C(\Omega)}+\Delta G(\Omega)\langle\hat{B}(\Omega)\rangle.
\end{split}
\end{equation}

The output measurement operator, including the influence of  a close feedback loop, can be written as
\begin{equation}\label{eq:measurement_data}
\begin{split}
\hat{y}(\Omega)&=\hat{x}(\Omega)+\hat{n}(\Omega)\\&=\frac{\chi_q(\Omega)\hbar\alpha\hat{a}_1(\Omega)+\chi_q(\Omega)F^{\rm zp}_{\rm th}(\Omega)+\hat{n}_{\rm zp}(\Omega)}{1+\chi_q(\Omega)C(\Omega)}\\&+\frac{\chi_m(\Omega)F_{\rm th}^{\rm cl}(\Omega)+n_{\rm cl}(\Omega)}{1+\chi_m(\Omega)C(\Omega)}+\Delta G(\Omega)\langle\hat{B}(\Omega)\rangle.
\end{split}
\end{equation}\\

It should be noted that the two denominators in the above equation are different, therefore the influence of the feedback kernel on the measurement operator can not be written in a factorized way and thereby can not be removed.  

\section{The effective Hamiltonian in the non-quadratic regime}
\label{appendix:non-quadrature}
\begin{widetext}

In this section, we consider the Schrödinger-Newton (SN) gravity in the non-quadratic regime, specifically when $\Delta X\gg\Delta x_{\rm int}$. In this scenario, the mass distribution of each atom is given by:
\begin{equation}
\rho_i(x_i,y_i,z_i)=\frac{1}{\sqrt{(2\pi)^3}\sigma_x\sigma_y\sigma_z}
{\rm exp}\left[-\frac{(x_i-\bar{x}_i)^2}{2\sigma_x^2}-\frac{(y_i-\bar{y}_i)^2}{2\sigma_y^2}-\frac{(z_i-\bar{z}_i)^2}{2\sigma_z^2}\right],
\end{equation}
where $(\bar{x}_i,\bar{y}_i,\bar{z}_i)$ represents the equilibrium position of the $i$-th atom. The self-gravity term is expressed as follows:
\begin{equation}
\begin{split}
\hat{H}_{\rm SN}(x_i,y_i,z_i)&=-\sum_{i,j}Gm_im_j\int\frac{\rho_i(x_i,y_i,z_i)}{|\hat{r}_i-r_j|}dr_i\\
&=-\sum_{i,j}\frac{Gm_im_j}{\sqrt{(2\pi)^3}\sigma_x\sigma_y\sigma_z}\int\frac{
{\rm exp}\left[-\frac{(x_i-\bar{x}_i)^2}{2\sigma_x^2}-\frac{(y_i-\bar{y}_i)^2}{2\sigma_y^2}-\frac{(z_i-\bar{z}_i)^2}{2\sigma_z^2}\right]}{\sqrt{(\hat{x}_i-x_j)^2+(\hat{y}_i-y_j)^2+(\hat{z}_i-z_j)^2}}dr_j.
\end{split}
\end{equation}

For simplification, we redefine the variables as:
\begin{equation}
\tilde{x}_j=\frac{x_j-\bar{x}_j}{\sigma_x},\quad\tilde{y}_j=\frac{y_j-\bar{y}_j}{\sigma_y},\quad\tilde{z}_j=\frac{z_j-\bar{z}_j}{\sigma_z},
\end{equation}
\begin{equation}\label{eq:integrand_variable}
\hat{\tilde{x}}_i=\frac{\hat{x}_i}{\sigma_x},\quad\hat{\tilde{y}}_i=\frac{\hat{y}_i}{\sigma_y},\quad\hat{\tilde{z}}_i=\frac{\hat{z}_i}{\sigma_z},\quad{\rm and}\quad
\tilde{c}_1=\frac{\sigma_x}{\sigma_y},\quad \tilde{c}_2=\frac{\sigma_x}{\sigma_z}.
\end{equation}
Then the SN Hamiltonian can be rewritten as:
\begin{equation}
\begin{split}
\hat{H}_{SN}&=-\sum_{i,j}\frac{Gm_im_j}{\sqrt{(2\pi)^3}\sigma_x}
\int\frac{{\rm exp}\left[-\frac{(\tilde{x}_j^2+\tilde{y}_j^2+\tilde{z}_j^2)}{2}\right]}{\sqrt{(\hat{\tilde{x}}_i-\tilde{x}_j-\frac{\bar{x}_j}{\sigma_x})^2+\frac{1}{\tilde{c}_1^2}(\hat{\tilde{y}}_i-\tilde{y}_j-\frac{\bar{y}_j}{\sigma_y})^2+\frac{1}{\tilde{c}_2^2}(\hat{\tilde{z}}_i-\tilde{z}_j-\frac{\bar{z}_j}{\sigma_z})^2}}d\tilde{x}_jd\tilde{y}_jd\tilde{z}_j,
\end{split}
\end{equation}
where integral is defined as:
\begin{equation}
\begin{split}
I_{ij}(\hat{\tilde{x}},\hat{\tilde{y}},\hat{\tilde{z}})=\int\frac{{\rm exp}\left[-\frac{(\tilde{x}_j^2+\tilde{y}_j^2+\tilde{z}_j^2)}{2}\right]}{\sqrt{(\hat{\tilde{x}}_i-\tilde{x}_j-\frac{\bar{x}_j}{\sigma_x})^2+\frac{1}{\tilde{c}_1^2}(\hat{\tilde{y}}_i-\tilde{y}_j-\frac{\bar{y}_j}{\sigma_y})^2+\frac{1}{\tilde{c}_2^2}(\hat{\tilde{z}}_i-\tilde{z}_j-\frac{\bar{z}_j}{\sigma_z})^2}}d\tilde{x}_jd\tilde{y}_jd\tilde{z}_j.
\end{split}
\end{equation}
To solve this integral, we reduce the three-dimensional problem to a one-dimensional problem by considering a scenario where $\sigma_x\gg\sigma_y$  and $\sigma_x\gg\sigma_z$ (or $\tilde{c}_1\gg1$ and $\tilde{c}_2\gg1$). When the lattice constant $a$ is substantially larger than $\sigma_x$, mutual gravity becomes negligible compared to self-gravity. Thus, we focus solely on the integration corresponding to self-gravity\,(i.e. $i=j$).
\begin{equation}\label{eq:intergration_1}
\begin{split}
I_{ii}(\hat{\tilde{x}},\hat{\tilde{y}},\hat{\tilde{z}})=\int\frac{{\rm exp}\left[-\frac{(\tilde{x}_j^2+\tilde{y}_j^2+\tilde{z}_j^2)}{2}\right]}{(\sqrt{2\pi})^3\sqrt{(\hat{\tilde{x}}_i-\tilde{x}_i)^2+\frac{1}{a^2}(\hat{\tilde{y}}_i-\tilde{y}_i)^2+\frac{1}{b^2}(\hat{\tilde{z}}_i-\tilde{z}_i)^2}}d\tilde{x}_id\tilde{y}_id\tilde{z}_i.
\end{split}
\end{equation}

Focusing on motion in the $x$-direction, with $\tilde{c}_1\gg1$ and $\tilde{c}_2\gg1$, we assume that quantum uncertainty is only contributed by the driving of the quantum radiation pressure noise, implying that the quantum uncertainty of the atom in the $y$- and $z$-directions is dominated by atom's motion around its crystal site $\delta y_i$ and $\delta z_i$. Using current experimental parameters, the ratio $\tilde{c}_1=\tilde{c}_2=\sigma_x/\sigma_{y/z}=\Delta X/\delta_{y/z}=5.47\times10^2$. As an estimate, we set $\hat{\tilde{y}}_i$ and $\hat{\tilde{z}}_i$ as random variables following a normal distribution with unit variance. The numerical results of Eq.\,\eqref{eq:intergration_1} are plotted in Fig.\,\ref{fig:numerical_integration}, which can be approximately fitted by a Gaussian function:
\begin{equation}
I(\hat{x})=A\,{\rm exp}\left[-\frac{x^2}{2b_1\sigma_x^2}\right].
\end{equation}

\begin{figure}
\includegraphics[width=1.0\textwidth]{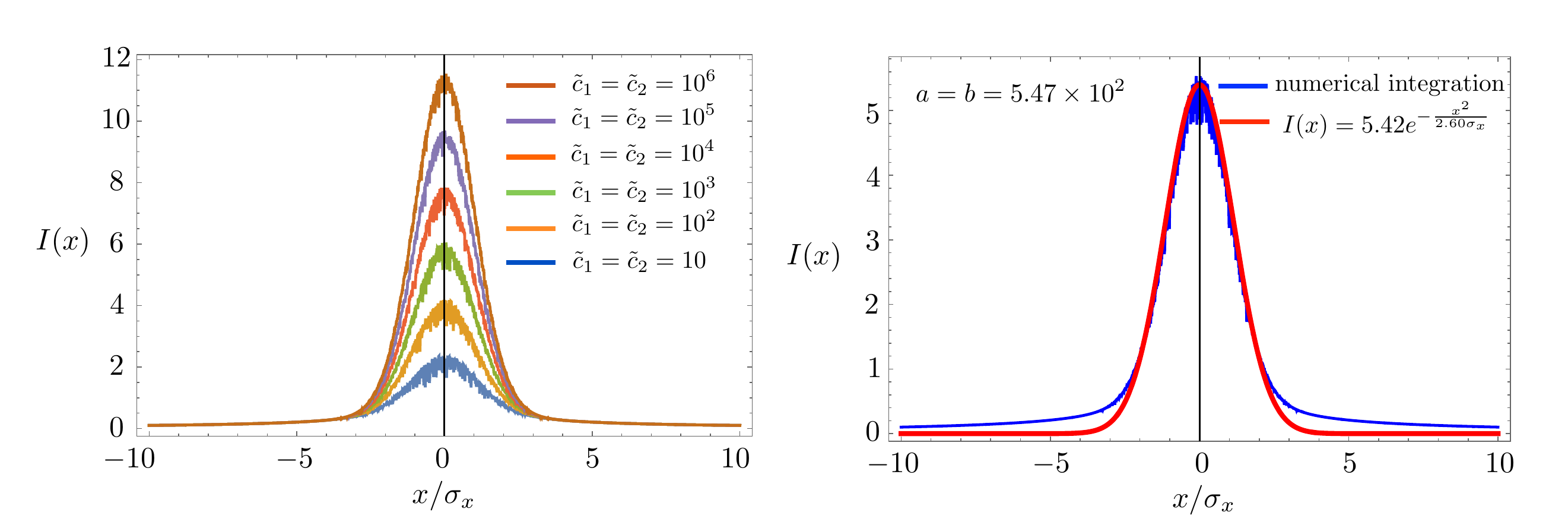}
\caption{The numerical results of the integration. The left panel shows the integration results for various ratios $\tilde{c}_1$ and $\tilde{c}_2$ in Eq.\,\eqref{eq:integrand_variable}, while the right panel depicts results for current experimental parameters. In the right panel, the blue line represents the numerical integration results, and the red line indicates the expression fitted by a Gaussian function.}\label{fig:numerical_integration}
\end{figure}

Finally, in the quasi-one-dimensional scenario, ignoring mutual gravity effects, the SN Hamiltonian simplifies to:
\begin{equation}
H_{\rm SN}(x)=-\frac{AGMm}{\sigma_x}{\rm exp}\left[-\frac{x^2}{2b_1\sigma^2_x}\right].
\end{equation} 

When we consider the case of  rotational motion, the SN Hamiltonian is modified as,
\begin{equation}
H_{\rm SN}(\Theta)=-\sum_i\frac{AGm^2}{r_i\sigma_{\Theta}}{\rm exp}\left[-\frac{\Theta^2}{2b_1\sigma^2_\Theta}\right]
=-\frac{AGMm}{\tilde{r}\sigma_{\Theta}}{\rm exp}\left[-\frac{\Theta^2}{2b_1\sigma^2_\Theta}\right],
\end{equation}
where $\tilde{r}=n/(\sum_i1/r_i)$ is a geometrical factor, where $n$ is the total number of atoms.

\section{The mutual gravity effect}
\label{appendix:mutual gravity}
In this section, we delve into the effects of mutual gravity, which becomes significant when the crystal constant $a$ is greater than or equal to $\sigma_x$.  The gravitational influence on the $i$-th atom due to SN gravity (encompassing both mutual and self-gravity) can be expressed as:
\begin{equation}
H^i_{SN}=-\frac{Gmm}{\sqrt{(2\pi)^3}\sigma_x}\sum_{j}\int\frac{{\rm exp}\left[-\frac{(\tilde{x}_j^2+\tilde{y}_j^2+\tilde{z}_j^2)}{2}\right]}{\sqrt{(\hat{\tilde{x}}_i-\tilde{x}_j-\frac{\bar{x}_j}{\sigma_x})^2+\frac{1}{\tilde{c}_1^2}(\hat{\tilde{y}}_i-\tilde{y}_j-\frac{\bar{y}_j}{\sigma_y})^2+\frac{1}{\tilde{c}_2^2}(\hat{\tilde{z}}_i-\tilde{z}_j-\frac{\bar{z}_j}{\sigma_z})^2}}d\tilde{x}_jd\tilde{y}_jd\tilde{z}_j.
\end{equation}

\begin{figure*}
\centering
\includegraphics[width=0.9\textwidth]{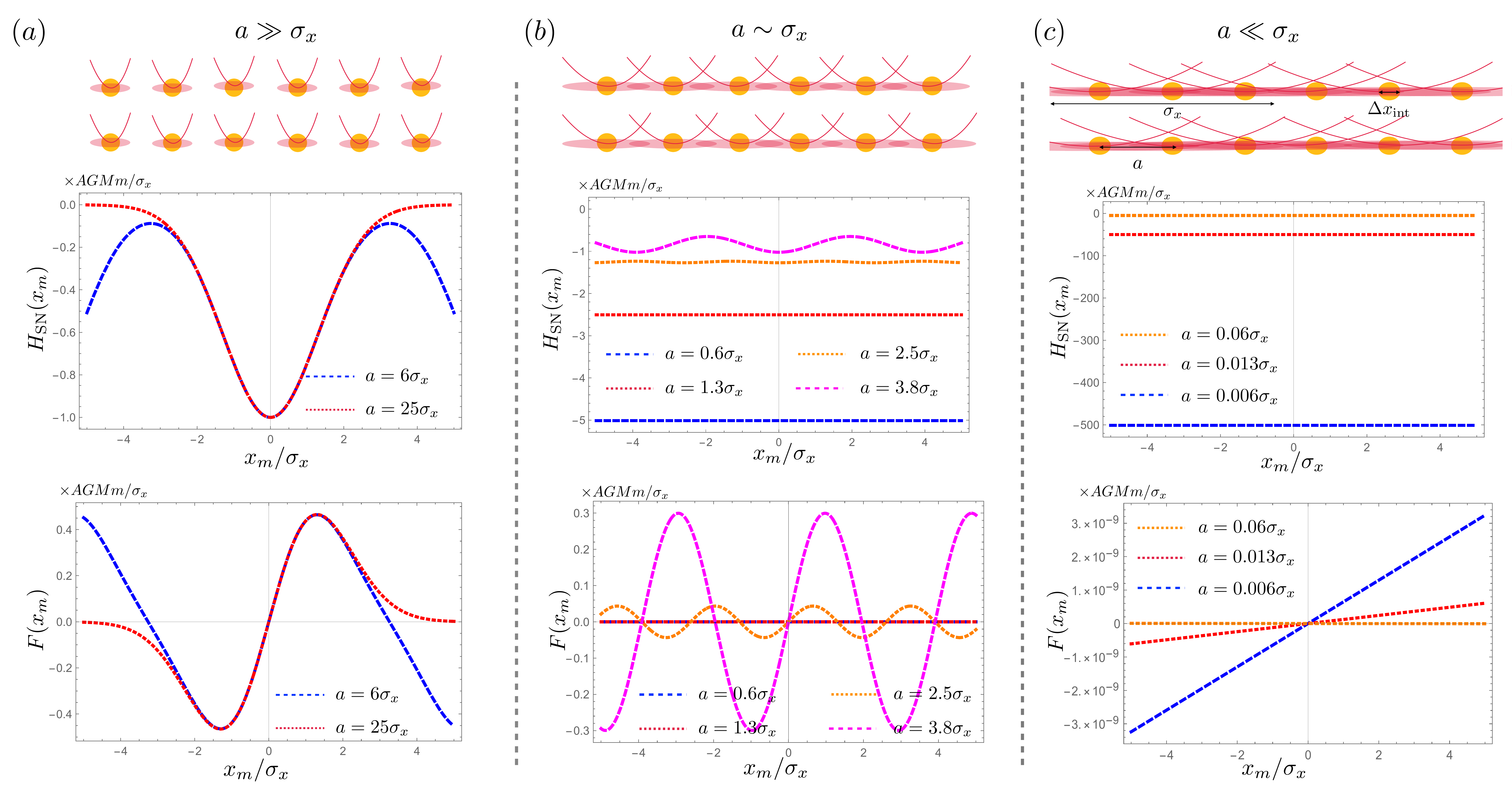}
\caption{SN gravity potential and SN gravity force considering mutual SN gravity. Panels (a), (b), and (c) illustrate the cases where $a\gg\sigma_x$, $a\sim\sigma_x$ and $a\ll\sigma_x$, respectively. As macroscopic uncertainty increases, mutual gravity interactions become progressively more significant, and the SN gravity potential becomes increasingly smooth. This trend indicates that the SN effect is gradually diminishing with the increased uncertainty of the RZ motion.}\label{fig:a_sigma_x}
\end{figure*}

Similar to the previous section, we utilize numerical integration techniques to demonstrate the SN interaction Hamiltonian experienced by a single atom, as depicted in the Fig.\,\ref{fig:a_sigma_x}. In the quasi-one-dimensional case, as the macroscopic uncertainty increases to a level comparable with or exceeding the lattice constant $a$, the SN Hamiltonian for a single atom becomes progressively less pronounced. This flattening indicates that SN mutual gravity increasingly counteracts SN self-gravity, thereby significantly weakening the SN effect. Therefore, in experiments aimed at detecting SN gravity, it is crucial to minimize the impact of SN mutual gravity.
\end{widetext}

\end{document}